\documentclass[11pt, a4paper]{article}

\usepackage[utf8]{inputenc}
\usepackage[T1]{fontenc}

\usepackage[margin=2.6cm]{geometry}
\usepackage{amsmath, amssymb}
\usepackage{graphicx}
\usepackage{booktabs}
\usepackage{array}
\usepackage{enumitem}
\usepackage{pdflscape}
\usepackage{capt-of}
\usepackage{diagbox}
\usepackage{placeins}
\usepackage{flafter}
\usepackage[british]{babel}
\usepackage{csquotes}
\usepackage[style=numeric-comp, sorting=none, maxbibnames=10, urldate=long, backend=biber]{biblatex}
\usepackage{microtype}
\usepackage{xcolor}
\usepackage{xurl}
\usepackage{hyperref}
\definecolor{linkblue}{rgb}{0.0, 0.2, 0.55}
\hypersetup{
    colorlinks=true,
    linkcolor=linkblue,
    citecolor=linkblue,
    urlcolor=linkblue,
    filecolor=linkblue,
    pdftitle={Portfolio Optimisation with Causal Networks},
    pdfauthor={Joshua Muthu}
}

\newcommand{\corrhrp}{\textsc{correlation-hrp}}
\newcommand{\skelhrp}{\textsc{skeleton-hrp}}
\newcommand{\undirhrp}{\textsc{undirected-hrp}}
\newcommand{\semcovhrp}{\textsc{semcov-hrp}}
\newcommand{\topohrp}{\textsc{topo-hrp}}
\newcommand{\toposemcovhrp}{\textsc{topo-semcov-hrp}}
\newcommand{\oriented}{\textsc{oriented}}
\newcommand{\lwhrp}{\textsc{lw-hrp}}
\newcommand{\dfhrp}{\textsc{defactored-hrp}}
\newcommand{\pcorrhrp}{\textsc{pcorr-hrp}}
\newcommand{\structrp}{\textsc{struct-rp}}
\newcommand{\coanchrp}{\textsc{coancestry-hrp}}
\newcommand{\corrherc}{\textsc{correlation-herc}}
\newcommand{\skelherc}{\textsc{skeleton-herc}}
\newcommand{\semcovherc}{\textsc{semcov-herc}}

\newcommand{\floatnote}[1]{\par\vspace{3pt}\begin{minipage}{0.92\textwidth}\footnotesize Note. #1\end{minipage}\par}

\appto{\bibsetup}{\sloppy}

\newcommand{\rsNtrials}{146}
\newcommand{\rsNtrialsPhaseI}{41}

\newcommand{\rsKurtosis}{15.9}
\newcommand{\rsVprimePSR}{0.953}

\newcommand{\rsCorrSharpe}{0.427}
\newcommand{\rsCorrSharpeWone}{0.411}
\newcommand{\rsCorrSharpeWthree}{0.407}
\newcommand{\rsCorrSharpeWfive}{0.383}
\newcommand{\rsCorrPSR}{0.964}
\newcommand{\rsCorrDSR}{0.925}

\newcommand{\rsDonePSR}{0.958}

\newcommand{\rsDlwSharpe}{0.394}
\newcommand{\rsDlwSharpeWone}{0.389}
\newcommand{\rsDlwSharpeWthree}{0.382}
\newcommand{\rsDlwSharpeWfive}{0.356}
\newcommand{\rsDdfSharpe}{0.409}
\newcommand{\rsDdfSharpeWone}{0.411}
\newcommand{\rsDdfSharpeWthree}{0.425}
\newcommand{\rsDdfSharpeWfive}{0.402}

\newcommand{\rsDpcSharpe}{0.410}
\newcommand{\rsDpcSharpeWone}{0.372}
\newcommand{\rsDpcSharpeWthree}{0.399}
\newcommand{\rsDpcSharpeWfive}{0.382}

\newcommand{\rsSpaDirWone}{0.361}
\newcommand{\rsSpaDirWtwo}{0.333}
\newcommand{\rsSpaDirWthree}{0.175}
\newcommand{\rsSpaDirWfive}{0.076}
\newcommand{\rsSpaDirWonePre}{0.217}
\newcommand{\rsSpaDirWtwoPre}{0.299}
\newcommand{\rsSpaDirWthreePre}{0.328}
\newcommand{\rsSpaDirWfivePre}{0.173}
\newcommand{\rsSpaSkelWone}{0.673}
\newcommand{\rsSpaSkelWtwo}{0.886}
\newcommand{\rsSpaSkelWthree}{0.273}
\newcommand{\rsSpaSkelWfive}{0.213}
\newcommand{\rsSpaSkelWonePre}{0.440}
\newcommand{\rsSpaSkelWtwoPre}{0.785}
\newcommand{\rsSpaSkelWthreePre}{0.346}
\newcommand{\rsSpaSkelWfivePre}{0.235}
\newcommand{\rsSpaSkelPooled}{0.827}
\newcommand{\rsSpaSkelPooledPre}{0.823}
\newcommand{\rsSpaDirPooled}{0.399}
\newcommand{\rsSpaDirPooledPre}{0.451}
\newcommand{\rsMcsSize}{16}
\newcommand{\rsMcsUniverse}{17}

\newcommand{\rsRewardSNR}{0.05}

\newcommand{\rsContrastSETotal}{0.019}
\newcommand{\rsContrastMDETotal}{0.047}
\newcommand{\rsContrastSEOrient}{0.015}
\newcommand{\rsContrastMDEOrient}{0.037}
\newcommand{\rsSeedN}{20}
\newcommand{\rsSeedMin}{0.375}
\newcommand{\rsSeedMedian}{0.389}
\newcommand{\rsSeedMax}{0.396}
\newcommand{\rsSeedRange}{0.021}

\newcommand{\rsOosOrientBothSeWfive}{0.087}
\newcommand{\rsOosOrientBothSeWone}{0.105}
\newcommand{\rsOosOrientBothSeWthree}{0.107}
\newcommand{\rsOosOrientBothSeWtwo}{0.107}
\newcommand{\rsOosOrientBothWfive}{+0.110}
\newcommand{\rsOosOrientBothWone}{+0.002}
\newcommand{\rsOosOrientBothWthree}{+0.079}
\newcommand{\rsOosOrientBothWtwo}{+0.013}
\newcommand{\rsOosOrientSeWfive}{0.065}
\newcommand{\rsOosOrientSeWone}{0.078}
\newcommand{\rsOosOrientSeWthree}{0.061}
\newcommand{\rsOosOrientSeWtwo}{0.086}
\newcommand{\rsOosOrientWfive}{+0.036}
\newcommand{\rsOosOrientWone}{+0.035}
\newcommand{\rsOosOrientWthree}{+0.105}
\newcommand{\rsOosOrientWtwo}{+0.050}
\newcommand{\rsOosResidSeWfive}{0.107}
\newcommand{\rsOosResidSeWone}{0.138}
\newcommand{\rsOosResidSeWthree}{0.110}
\newcommand{\rsOosResidSeWtwo}{0.116}
\newcommand{\rsOosResidWfive}{-0.001}
\newcommand{\rsOosResidWone}{-0.090}
\newcommand{\rsOosResidWthree}{+0.005}
\newcommand{\rsOosResidWtwo}{-0.001}
\newcommand{\rsOosSkelSeWfive}{0.056}
\newcommand{\rsOosSkelSeWone}{0.076}
\newcommand{\rsOosSkelSeWthree}{0.065}
\newcommand{\rsOosSkelSeWtwo}{0.072}
\newcommand{\rsOosSkelWfive}{+0.100}
\newcommand{\rsOosSkelWone}{+0.107}
\newcommand{\rsOosSkelWthree}{-0.009}
\newcommand{\rsOosSkelWtwo}{+0.071}
\newcommand{\rsOosTotalSeWfive}{0.076}
\newcommand{\rsOosTotalSeWone}{0.122}
\newcommand{\rsOosTotalSeWthree}{0.077}
\newcommand{\rsOosTotalSeWtwo}{0.122}
\newcommand{\rsOosTotalWfive}{+0.136}
\newcommand{\rsOosTotalWone}{+0.142}
\newcommand{\rsOosTotalWthree}{+0.096}
\newcommand{\rsOosTotalWtwo}{+0.122}

\title{Portfolio Optimisation with Causal Networks}
\author{Joshua Muthu\\[2pt]
  \normalsize Independent researcher, London\thanks{This work was
  carried out while the author was at the Department of Computing,
  Imperial College London.}\\
  \normalsize\href{mailto:joshuasemuthu@gmail.com}{joshuasemuthu@gmail.com}}
\date{}

\begin{document}

\maketitle

\begin{abstract}
\noindent Hierarchical portfolio allocators typically take the return correlation matrix as their input, but correlation carries no notion of direction or causality. Causal discovery algorithms supply this as a directed acyclic graph (DAG), which can be decomposed into a skeleton (which assets are linked) and edge orientations (which assets affect which). This paper asks whether replacing the correlation matrix in hierarchical allocation with a causal graph improves performance, and how much of any improvement comes from the skeleton rather than the orientations. I feed an identical DAG to two classes of allocators: one that uses only the graph's skeleton, and a new family that also uses its edge orientations. Comparing them with each other and with standard Hierarchical Risk Parity (HRP) decomposes the effect of the graph into its two components. I estimate this decomposition on a 99-asset subset of the S\&P~500 over 2007--2024, with four estimation windows from 9 months to 2 years, using DYNOTEARS and VARLiNGAM for causal discovery. Replacing the correlation matrix with a causal DAG does not improve the Sharpe ratio. Standard HRP outperforms the graph-fed allocators at the two shorter windows, while at the two longer windows, that is not the case (though the performance difference is slight). The skeleton lowers performance at every window and the edge orientations offset this, so the total effect is indistinguishable from zero. The orientations' contribution comes from the DAG's removal of the common market factor: a de-factored covariance built without any graph reproduces it. No comparison is statistically significant after accounting for the search over allocators.

\end{abstract}

\section{Introduction}
\label{chap:intro}

\subsection{Background and motivation}
\label{sec:motivation}

Portfolio optimisation allocates capital to different
assets in order to maximise returns for a given level of 
risk. Since Markowitz's seminal mean-variance model
\cite{markowitz1952portfolio}, the covariance
(or equivalently, correlation) matrix of returns has
been the main statistical object used for optimisation.
Hierarchical Risk Parity (HRP) is a recent
method \cite{lopezdeprado2016hrp} that extends the 
mean-variance literature and aims to address
the matrix instability problems that affected earlier 
models. HRP first clusters assets based on their 
correlational distance, producing a dendrogram (binary
tree), then recursively bisects down that tree and 
allocates capital to clusters inversely proportional
to their estimated variances.

However, correlation is an undirected metric: 
it records which assets co-move, but not which assets
move others \cite{lopezdeprado2023causal}. Crucially,
it lacks causal structure, so is prone to spurious
correlation \cite{simon1954spurious}.
Causal discovery algorithms provide a directed alternative
to the correlation matrix. Score-based structure learning algorithms
(such as NOTEARS \cite{zheng2018notears} and its temporal extension
DYNOTEARS \cite{pamfil2020dynotears}) can fit directed acyclic graphs
(DAGs) that can represent which assets are linked and the direction
of influence between them. Some papers have applied these DAGs to
equity markets \cite{howard2025causal,tang2024trading}, 
but do not use them for weighting or allocation within 
hierarchical portfolio allocators.

However, current hierarchical methods like HRP are not
able to represent directed objects like DAGs,
because HRP's clustering methodology requires symmetric matrices.
But a DAG can be separated into two information sources:
its skeleton, an undirected structure that displays which asset
pairs are causally linked; and its edge orientations, which indicate
the direction of each causal link. The skeleton can be directly
provided as an input to hierarchical allocators through any
orientation-blind transformation of the graph's adjacency matrix,
such as symmetrising it. The edge orientations cannot, as any such
transformation discards directionality,
so incorporating their information requires novel allocators.

\subsection{Research question}
\label{sec:question}

This paper addresses a central research question:

\begin{quote}
When the
correlation matrix in hierarchical portfolio allocation is
replaced by a causal graph, is there a performance
improvement, and how much of any improvement is attributable
to the graph's skeleton versus its edge orientations?
\end{quote}

\noindent I briefly define the terms in the question here, and
expand on them in more detail later on in the paper.
First, hierarchical portfolio allocation specifically refers to
HRP, where capital is invested fully and long-only. I invest over
a fixed 99-asset subset of the S\&P~500, with the portfolio
rebalanced monthly between 2007 and 2024. The asset ``universe''
is discussed in further detail in Section~\ref{chap:discovery},
the allocation protocol in Section~\ref{chap:evaluation}.
Second, the correlation matrix is the matrix originally used in HRP,
the estimation window's sample correlation
\cite{lopezdeprado2016hrp}. This feeds into the baseline HRP
control (see Section~\ref{chap:allocators}).
Third, a discovered causal graph is the output of one of the causal
discovery algorithms (DYNOTEARS, VARLiNGAM, with ridge-regularised
Granger as a control on edge magnitudes). It is a contemporaneous (no lags)
asset-to-asset graph that is re-fitted at each rebalance
at 4 different estimation windows 
(189, 252, 378 and 504 trading days; roughly 9 months to 2 years).
See Section~\ref{chap:discovery} for more details.
Fourth, performance improvement is defined as the out-of-sample 
annualised Sharpe ratio of daily portfolio returns, net of 
investment cost. I also use several other metrics that account 
for multiplicity (see Section~\ref{chap:evaluation}), including the 
Deflated Sharpe ratio, Hansen's superior predictive ability (SPA) test, and the Model Confidence Set
\cite{bailey2014deflated,hansen2005spa,hansen2011mcs}.
Finally, I compare the relative performance effect attributable to
the graph's skeleton versus its edge orientations using an ablation design, where an identical graph $M$ feeds
a variety of allocators that differ only in the information they
extract from the graph. The skeleton-only allocator, \skelhrp{},
clusters on an orientation-blind embedding of $M$, while three 
edge-orientation-aware allocators consume the directed matrix itself
(either through the allocation step, the ordering step, or both).
See Section~\ref{chap:allocators} for further details.

The question has two parts. The first is whether the causal graph
improves on the correlation matrix at all. I answer this in
Section~\ref{sec:graph-vs-correlation} by comparing the graph-fed allocators with
standard sample-correlation HRP. The second is the decomposition,
which requires an allocator that is able to
consume orientation information, which symmetric clustering cannot.
In Section~\ref{chap:allocators}, I construct the edge-orientation-aware family
and discuss the orientation-blindness of the other allocators. 

With that requirement met, I am able
to isolate the performance effect of the two components of the graph.
The skeleton's contribution over the HRP baseline measures the
value of knowing which assets are causally linked to each other. 
The edge orientations' contribution over the skeleton measures
the additional value of knowing the direction of the causal links
between assets, holding the graph fixed. The two marginal
contributions sum to the total performance effect of the causal
graph over the correlation-based allocator. 

In addition, after isolating the contributions of the two 
aspects of the graph, further questions arise regarding what drives the
performance contribution of each of those two aspects. Specifically:
\begin{itemize}
    \item Does the skeleton's performance contribution arise purely from the 
    actual discovered causal graph structure, or can that contribution
    be reproduced by any sparse dependence structure of a similar density,
    regardless of whether it has been created by causal discovery?
    \item Does the performance contribution from edge orientations
    arise specifically from the directional content of the causal DAG's
    edges, or does it arise from another aspect of the DAG, such as the properties of the
    implied covariance, which can be reproduced without the need for
    a causal graph?
\end{itemize}
\noindent I answer these questions in Section~\ref{chap:evaluation}.

\subsection{Contributions and main findings}
\label{sec:challenges}

There are three key technical challenges I must address in this paper.
They bring together three research domains,
which have previously not been applied jointly to portfolio allocation:
causal structure learning; hierarchical portfolio construction with
directed, asymmetric inputs; and rigorous backtesting. The three
challenges are as follows:

\begin{itemize}[itemsep=2pt, topsep=3pt]
  \item Correct attribution of performance effects to different aspects of 
  the causal graph (Section~\ref{sec:ablation-design} and Section~\ref{chap:evaluation}). Naively replacing the
  correlation matrix with a causal graph means that different aspects
  of the graph (e.g. its skeleton, its edge orientations, and its
  implied covariance) all inform allocation jointly. This means
  the contributions of specific aspects cannot be individually isolated.
  I also need to identify whether the performance effects I attribute to
  the skeleton and the edge orientations of the graph are actually products
  of the causal graph, and are not just byproducts from other, non-graph-specific mechanisms that the graph just happens to possess
  (such as generic shrinkage or market-factor removal).
  No prior work has performed this isolation; it requires a credible,
  reproducible graph, which Section~\ref{chap:discovery} supplies. 
  \item Getting hierarchical portfolio allocators to accept 
  asymmetric DAGs as
  an input (Section~\ref{chap:allocators}). Hierarchical allocators
  use symmetric distance matrices as inputs. Simply symmetrising
  a DAG's adjacency matrix would discard the edge orientation information, but this paper's research question needs to isolate orientation's effect on performance. There is currently no established way to feed
  an asymmetric DAG adjacency matrix into the clustering or allocation step of these allocators.
  \item Performing statistical inference at realistic effect sizes
    (Section~\ref{chap:evaluation}). The main metric for the ``performance improvement'' in the paper's research question is the
    difference in net Sharpe between allocators. These differences are
    likely to be small relative to standard error, given the 18 years of
    monthly rebalances and the heavy-tailed nature of returns. Also, this
    study evaluates \rsNtrials\ configurations (across different discovery methods, allocators, and window sizes). Deciding whether any
    difference in net Sharpe is not just an artefact
    of the search over many configurations requires statistical inference
    that accounts for this search, and an estimate of the statistical power of the tests I use. 
\end{itemize}

\noindent The three challenges above are respectively addressed by the three
contributions below. The novelty lies in the first two; the third applies
established inference tools to the question:

\begin{itemize}[itemsep=2pt, topsep=3pt]
  \item An ablation design with mechanism controls, plus a reproducible
    discovery pipeline (Section~\ref{sec:ablation-design}, Section~\ref{chap:evaluation} and Section~\ref{chap:discovery}). 
    The same per-window DYNOTEARS or VARLiNGAM graph feeds several different
    allocators that incrementally take in more information from the DAG,
    from none (\corrhrp{}), to skeleton-only, to edge-orientation-aware.
    This means that each increment isolates one aspect of the graph:
    the skeleton's contribution is the skeleton-only allocator minus \corrhrp{}, and the
    orientations' contribution is the edge-orientation-aware allocator minus the
    skeleton-only one. 
    I ensure that the allocators receive the identical graph and are replicable.
    Three graph-free controls then test whether each contribution is a product of the graph. A Ledoit--Wolf shrunk covariance and a covariance with the common market factor removed each change the covariance in a way similar to the DAG, without any graph, and a partial-correlation network of comparable density to the skeleton tests whether any sparse network reproduces the skeleton's contribution.
  \item Creating a new family of edge-orientation-aware allocators
    (Section~\ref{chap:allocators}). This is the core methodological novelty
    of my research. I provide HRP with two directed inputs, one per step of the
    algorithm. At the allocation step, the sample covariance is replaced by
    the covariance implied by the DAG's structural equation model (SEM).
    At the clustering (ordering) step, HRP's clustering dendrogram is replaced by
    the DAG's topological order. \semcovhrp{} uses the
    implied covariance at the allocation step, \topohrp{} uses the topological ordering at the ordering step, while
    \toposemcovhrp{} uses both the topological ordering and the implied covariance. Two further edge-orientation-aware allocators are placed with every other allocator in Table~\ref{tab:allocator-map}.
  \item Evaluating the allocators' performance at realistic effect sizes (Section~\ref{chap:evaluation}). Each comparison is tested with
    a stationary block bootstrap \cite{politis1994stationary}, so that the
    resulting $p$-values account for autocorrelation and heavy tails.
    To account for all \rsNtrials\ configurations evaluated, I use the Deflated Sharpe
    ratio, which corrects the best-performing configuration based on the number of trials examined. I also use Hansen's SPA test and the Model Confidence Set
    to identify whether any member of a set of strategies outperforms the baseline once the whole set is tested jointly, including a pooled test over both strategies and
    estimation windows. 
    I also compute the minimum detectable effect at $80\%$ power
    (about \rsContrastMDEOrient{} Sharpe for the edge orientation and
    \rsContrastMDETotal{} for the total), so that any null result
    acts as an upper bound on the effect rather than as evidence of no effect at all.
    I then test on 2025--2026 out-of-sample data and, as a robustness check, I rerun the decomposition's three anchor allocators under a second hierarchical allocator, Hierarchical Equal Risk Contribution (HERC).
\end{itemize}

\label{sec:quantified-results}

I find that the causal graph does not yield a performance improvement over the correlation matrix as an input to hierarchical allocation. The graph's skeleton has a negative effect on performance at every estimation window, while the edge orientations have a positive effect relative to the skeleton. The edge orientations' relative performance improvement is due to the fact that the implied covariance removes the common market factor, so it does not actually require a graph. I find the total net effect of using the skeleton and the edge orientations together (relative to the correlational baseline) to be statistically indistinguishable from zero at every estimation window. None of the comparisons between different allocators are statistically significant after accounting for the search over allocators.
The headline results are as follows, with the full results discussed in
Section~\ref{chap:evaluation}:

\begin{itemize}[itemsep=2pt, topsep=3pt]
  \item The causal graph yields no performance gain over correlation. \semcovhrp{}, the allocator that uses the DAG's implied covariance for allocation, has a difference of $-0.017$ to $+0.013$ Sharpe (net of 5\,bps transaction costs) compared to \corrhrp{} (the baseline HRP allocator) across the 4 estimation windows. None of those results are significant (all $p \geq 0.36$). 
  Hansen's SPA test, which is conducted across all the causal allocators against \corrhrp{}, does not reject the null hypothesis of no gain at any estimation window, nor when the windows are pooled.
  \item Decomposing the causal graph's performance into its two aspects shows that both have consistent relative effects. At every window, the skeleton has a negative effect ($-0.003$ to $-0.028$), while the edge orientations have a positive effect ($+0.009$ to $+0.021$). The two aspects' effects roughly offset each other so that the total is indistinguishable from zero.
  \item A generic, graph-free sparse network replicates the performance impact from the discovered graph's skeleton: a partial-correlation network of comparable density, built without any causal discovery, matches the net Sharpe of the skeleton-only allocators at every window (the differences range from $-0.020$ to $+0.012$, $p \geq 0.07$), with both underperforming \corrhrp{}. This answers the first sub-question of Section~\ref{sec:question}: the skeleton's contribution can be reproduced by any sparse dependence structure of similar density.
  \item The performance gain attributable to the edge orientations is due to market-factor removal, not edge directions: a de-factored covariance, built without any graph, reproduces the net Sharpe of the edge-orientation-aware allocators at every window (with the differences between them ranging from $-0.011$ to $+0.001$). The orientation gains are below the minimum detectable effect, which bounds the edge-orientation-aware allocators to yielding a maximum gain of approximately $0.04$ Sharpe over the skeleton-only allocators. Note that the above results are specific to DYNOTEARS-discovered graphs; under VARLiNGAM graphs, they show no systematic gain (6 of 12 contrasts positive, mean $0.000$). This answers the second sub-question.
\end{itemize}

On the 2025--26 out-of-sample slice (consisting of 19 rebalances, from January 2025 to August 2026), I find that the marginal performance impacts of the orientation contribution and the total were positive at every window: see Section~\ref{sec:oos}.

\label{sec:structure}

I conduct a literature review in Section~\ref{chap:background}. 
In Section~\ref{chap:discovery}, I discuss the data I use, as well as the causal discovery algorithms that produce the graphs necessary for this research.
Then Section~\ref{chap:allocators} defines the novel allocators I create and sets out the ablation design.
Section~\ref{chap:evaluation} then conducts the decomposition and statistical inference to answer the research question.
Section~\ref{chap:conclusion} concludes.
The full result tables, hyperparameters, and reproduction instructions are located in Appendix~\ref{app:tables}.

\section{Background and related work}
\label{chap:background}

This paper is underpinned by the three research domains identified in Section~\ref{sec:challenges}: hierarchical portfolio allocation, together with the literature on rigorous backtesting that has grown up around it (Section~\ref{sec:hierarchical-allocation}); applying causal graphs to investing (Section~\ref{sec:causal-investing}); and causal discovery (Section~\ref{sec:causal-discovery}). 

\subsection{Hierarchical portfolio allocation}
\label{sec:hierarchical-allocation}

In order to weight assets in a portfolio to minimise risk (variance) while maximising return, mean-variance optimisation \cite{markowitz1952portfolio} requires inverting the covariance matrix of asset returns.
For a large set of assets at a standard window size (around 1 year), that matrix is ill-conditioned, which means that the resulting weights are highly concentrated and unstable.
Hierarchical Risk Parity \cite{lopezdeprado2016hrp} sidesteps this matrix inversion problem. It uses a three-step approach to optimisation. First, it clusters assets by converting the
correlation matrix $\rho$ into a distance matrix $d$, where $d_{ij} = \sqrt{\tfrac{1}{2}(1-\rho_{ij})}$, and builds a binary tree (dendrogram) so closely related assets cluster together.
The second step is quasi-diagonalisation, where assets are ordered in the dendrogram's leaf order, so that large covariances sit near the diagonal.
Finally, assets are weighted in the portfolio using recursive bisection: the ordered list is split in half, the two halves are then weighted in inverse proportion to their cluster variances,
then each half of the list is recursed into. 
HRP is robust to ill-conditioning as no matrix is inverted.
This yields performance improvements over mean-variance and inverse-variance portfolios out of sample \cite{lopezdeprado2016hrp}.

Note that the allocation step (recursive bisection) takes in an ordered list, plus the variances from the clustering step. In standard HRP, that ordered list comes from a dendrogram. However, this need not be the case: any procedure which produces
an economically meaningful ordering of the assets can be used in place of the clustering step.
In Section~\ref{sec:orientation-family}, I do so using a causal-ordered bisection.
Also, as stated in Sections~\ref{sec:motivation} and \ref{sec:challenges}, HRP clustering requires symmetric matrices, which means
it cannot represent directed causal influence.
Section~\ref{chap:allocators} addresses the problem of augmenting HRP with directional information.

There are several variants of HRP, including hierarchical
clustering-based asset allocation \cite{raffinot2017hierarchical}, nested clustered optimisation
\cite{lopezdeprado2020mlam} and multi-asset, factor-based HRP \cite{lohre2020hrp}.

Hierarchical Equal Risk Contribution (HERC) \cite{raffinot2018herc} is a further hierarchical allocator: rather than recursively bisecting the leaf order of the dendrogram, it equalises risk contributions across the clusters the dendrogram identifies. I use it as a robustness check on the decomposition in Section~\ref{sec:robustness}.

Hierarchical Sensitivity Parity (HSP) \cite{rodriguez2023hsp} keeps the HRP mechanism but clusters assets on the similarity of their sensitivities to a set of macro-financial drivers, estimated by a feed-forward neural network per asset, rather than on return correlation. It is not a causal allocator, since the drivers are selected by their correlation with the assets and the clustering input is still a symmetric distance, and because the sensitivities come from randomly initialised networks the allocation is not exactly reproducible. It was the baseline of the first phase of this study (Appendix~\ref{app:hsp}).

Aside from simply improving the allocator mechanism as a whole, another way to improve portfolio optimisation is to improve the input. Linear shrinkage can reduce noise in the sample covariance matrix by punishing deviation from a structured target like an identity matrix, trading off a small bias for a large reduction in variance \cite{ledoitwolf2004}.
Another method for de-noising assets' co-movement is to use a single-factor model \cite{sharpe1963simplified} to remove the factor common to all assets' returns, and then use the resulting residual (de-factored) covariance.
The literature usually identifies this ``common market factor'' with the largest eigenvalue of the correlation matrix \cite{laloux1999noise}. I use the simpler equal-weight cross-sectional mean return as the market proxy, regressing each asset on it and taking the covariance of the residuals.
Both of these methods de-noise the input to the allocator without the need for any DAG. 
I use both as controls in Section~\ref{sec:orientation-role} to identify whether the effect of the edge-orientation-aware allocators is due to the DAG itself or to these de-noising estimators.

In the same section, I similarly use a graph-free simulacrum of the skeleton-only allocator to identify the drivers of its effect. This graph-free control is a ``thresholded partial-correlation network''. 
The inverse of the covariance matrix (the precision matrix) gives the conditional dependence between pairs of assets. Graphical lasso \cite{friedman2008glasso} is the usual sparse estimator of this matrix. I take a simpler route: I invert a Ledoit--Wolf covariance, convert it to partial correlations, and keep the largest-magnitude entries up to the DAG's edge count. Either way the result is a sparse, undirected network with no need for causal discovery.

L\'opez de Prado and co-authors have also studied overfitting to backtests \cite{bailey2014deflated,lopezdeprado2023causal}: the problem of establishing that the performance of a chosen allocator configuration has not simply been selected out of the many configurations tried. This literature has produced metrics that account for the search.
The Probabilistic Sharpe ratio \cite{bailey2012psr} replaces the
Gaussian sampling theory of the Sharpe ratio, which is untenable at the
excess kurtosis of ${\approx}\,\rsKurtosis$ I measure here, with a
version that corrects for the sample's skewness, kurtosis and length.
The Deflated Sharpe ratio \cite{bailey2014deflated} then raises the
hurdle from zero to the best Sharpe expected from $N$ independent
trials under the null of no skill, which discounts selection
bias directly. For comparing a family of strategies against a baseline, White's
Reality Check \cite{white2000reality} and Hansen's SPA test
\cite{hansen2005spa} bootstrap the null hypothesis that no candidate in
the family beats the baseline, and the Model Confidence Set
\cite{hansen2011mcs} returns the subset of strategies that are
statistically indistinguishable from the best. I apply this battery to
the headline contrasts in Section~\ref{chap:evaluation}. The Deflated Sharpe ratio
corrects each headline Sharpe ratio for all \rsNtrials\ configurations I
evaluated, while the SPA and MCS run over the smaller family of
candidates relevant to each question. Several effects that look
important at first do not survive it, and I report those as well.

\subsection{Applying causal graphs to investing}
\label{sec:causal-investing}

The closest prior work to this paper is Howard, Lohre and Mudde's application of DYNOTEARS to S\&P~500 equities, re-estimated on a rolling basis \cite{howard2025causal}.
They find that the resulting networks carry regime information that correlation alone does not, and that causal networks outperform correlation-based methods in some regards but not consistently. They also report two other findings that are relevant for this research.
First, the contemporaneous matrix from the structural model carries almost all of the structure. I therefore fit only one lag in Section~\ref{sec:discovery-run}, and just use the contemporaneous matrix as an input to my allocators.
Second, they do not use the graph as a direct input to portfolio allocation. Instead, they use it to represent similarity, or compress it into a single centrality metric. Whether a graph used directly as the allocation input improves on correlation, and whether any effect comes from the skeleton or from the edge orientations, is the gap this paper addresses.

\subsection{Causal discovery}
\label{sec:causal-discovery}

There are two main types of causal discovery algorithms: score-based and constraint-based. Constraint-based algorithms, such as PC (Peter--Clark) and fast causal inference (FCI), scale poorly as dimensionality increases, and rely on the faithfulness assumption (that statistical independence identified in the sample reflects d-separation in the true graph). This assumption is weak in real-world markets, as counterbalancing effects can yield a negligible correlation between assets despite a true causal connection existing \cite{howard2025causal}. The time-series constraint-based methods PCMCI and PCMCI+ \cite{runge2019pcmci,runge2020pcmciplus} are also unsuitable here, since the first tests only lagged links and the second returns a partially directed graph.
Score-based algorithms offer a better alternative. One example is NOTEARS \cite{zheng2018notears}, which turns graph discovery into a continuous optimisation problem by recasting the acyclicity requirement as a smooth constraint. DYNOTEARS \cite{pamfil2020dynotears} extends this to a structural vector autoregression (SVAR) representation for time series data, and is the primary causal discovery algorithm I use.

However, NOTEARS methods do have a limitation regarding edge orientations: the fact that the least-squares score they use to fit the DAG implicitly assumes that variables' shocks have equal variance. This assumption means NOTEARS tends to orient directed edges from lower-variance variables to higher-variance ones \cite{peters2014identifiability}. On that basis, Reisach, Seiler and Weichwald \cite{reisach2021beware} find that simply regressing each variable on lower-variance variables yields graphs that are approximately as good as NOTEARS' ones. Once the variables are standardised (each variable has its mean subtracted and is divided by its standard deviation, so it has mean 0 and variance 1), NOTEARS' accuracy falls to the level of PC. As a result of the above, I standardise the variables before fitting the discovery methods (see Section~\ref{sec:discovery-run}). Another paper by Reisach et al.\ \cite{reisach2023r2} also shows that, even after standardisation, ordering by explained variance ($R^2$) survives in NOTEARS graphs; I discuss this in Section~\ref{sec:caveats}.

Another causal discovery family is LiNGAM \cite{shimizu2006lingam}, which uses non-Gaussian residuals to identify causal orderings for DAGs. The identifying assumption is that, if $x$ causes $y$, then regressing $y$ on $x$ will result in a residual that is independent of $x$, while regressing $x$ on $y$ will leave a residual that is dependent on $y$. In this way, the causal parent can be identified by checking its independence against the residual. However, this requires the data to be non-Gaussian, as if the variables are Gaussian, both regressions will yield independent residuals. DirectLiNGAM \cite{shimizu11directlingam}, the estimator I use, iteratively regresses variables against each other and then removes the causal parent until the ordering is complete.
VARLiNGAM \cite{hyvarinen2010varlingam} is its structural-VAR form, and serves as my robustness comparator to DYNOTEARS. Its identification rests on heavy-tailed residuals, which financial returns exhibit in the short run. However, at longer return horizons, this is not the case, as aggregation pulls residuals towards Gaussianity \cite{cont2001stylized}.

The final causal discovery algorithm I use is a VAR(1) coefficient matrix that has been ridge-regularised, which yields a Granger-causal
\cite{granger1969investigating} directed graph. 
This gives a reasonable directional ordering, 
but its shrunken one-lag coefficients are biased. 
As a result, I use it as a control on edge magnitudes, separating the two aspects of orientation:
the directionality itself, and the magnitudes of the edges.

\subsection{Research gap}
\label{sec:gap}

Table~\ref{tab:gap} positions my research against prior work. HRP forms the baseline, \corrhrp{}, of my research. As Section~\ref{sec:hierarchical-allocation} noted, HRP does not use directionality in the clustering step. HSP clusters on driver sensitivities rather than correlation, but selects the drivers themselves by correlation, and still uses a symmetric distance matrix as an input to the allocator. Howard et al.\ use causal discovery to identify DAGs but do not use them directly for portfolio allocation. None of them can separately attribute the value of the skeleton and the edge orientation of a graph, as none of them set hierarchical portfolio weights from a causal graph of assets in the first place. I do so in Sections~\ref{chap:allocators} and~\ref{chap:evaluation}.

Allocating from a graph of assets is not itself novel: Mantegna \cite{mantegna1999hierarchical} creates a minimum spanning tree (MST) from the correlation matrix, and uses that to create a hierarchical clustering of assets. Tumminello et al.\
\cite{tumminello2005pmfg} extend this by using a planar maximally
filtered graph (PMFG), which keeps more edges.
Subsequent research along these lines found that allocating portfolio weight to the periphery of the network outperforms allocating weight to central nodes \cite{pozzi2013spread,peralta2016network}, as the latter are connected to everything and 
so diversify least. 
This research motivates the importance of the skeleton: \skelhrp{}, the skeleton-only
allocator of this paper, likewise builds a graph over the assets and allocates down its
structure, but its graph comes from causal discovery rather than from
filtering a correlation matrix. 
Directed-graph allocation is also
not unprecedented: prior research has set portfolio weights directly
from directed Granger-causality networks \cite{vyrost2019network}.
However, no prior work sets hierarchical portfolio weights from a discovered score-based causal graph, nor decomposes the allocation value of such a
graph into its skeleton and its edge orientations. This paper addresses that gap.

\begin{table}[htbp]
  \centering
  \caption{This study compared to prior work.}
  \label{tab:gap}
  \small
  \setlength{\tabcolsep}{4pt}
  \begin{tabular}{l >{\raggedright\arraybackslash}p{2.4cm} *{4}{>{\centering\arraybackslash}p{2.0cm}}}
    \toprule
    Approach & Clusters on & Causal discovery? & Weights from graph? & Asymmetric allocation? & Skeleton vs orientation? \\
    \midrule
    HRP \cite{lopezdeprado2016hrp} & correlation & no & n/a & no & n/a \\
    HSP \cite{rodriguez2023hsp} & driver sensitivities & no & no & no & no \\
    MST/PMFG \cite{pozzi2013spread,peralta2016network} & correlation network & no & yes & no & n/a \\
    Granger networks \cite{vyrost2019network} & (no clustering) & no & yes & yes & no \\
    Howard et al.\ \cite{howard2025causal} & (no allocation)  & yes & no  & no & no \\
    This study                     & causal graph & yes & yes & yes & yes \\
    \bottomrule
  \end{tabular}
  \floatnote{HRP is the baseline of this study, denoted \corrhrp{}.}
\end{table}

\section{Discovering the graph}
\label{chap:discovery}

This section provides the graph required for carrying out the ablation design (see Sections~\ref{sec:ablation-design} and~\ref{chap:evaluation}). This graph feeds into every allocator (except the baseline) at every monthly rebalance.
Section~\ref{sec:data} describes the data,
Section~\ref{sec:discovery-run} the rolling discovery and the
directional prior, Section~\ref{sec:asset-graph} the extracted graph, and finally Section~\ref{sec:comparators}
describes the two control graphs.

\subsection{Data}
\label{sec:data}

Given my compute constraints, the full S\&P~500 is too large to run causal discovery over.
The S\&P~100 was the natural target, but its historical membership is paywalled. I therefore took the union of the largest constituent assets of the S\&P~500 by market cap at three dates to approximate it. These three dates were the start, midpoint and end of the pilot runs that I conducted over 2018--2020, prior to running the full backtest. The resulting union consisted of 99 assets (listed in Appendix~\ref{app:universe}) which I held fixed across the whole 2007--2024 sample.
One drawback of this method of picking assets is that they were only picked from a narrow 2-year window: large firms that failed or shrank before then, such as Lehman Brothers and Bear Stearns, are excluded, while firms that listed after 2007, such as Meta, are included.
This introduces survivorship bias which raises the absolute performance levels of all the allocators I work with. However, the pairwise comparisons (which are just subtractions) are unaffected.
Quantifying the extent of survivorship bias is left to future work.
I use CRSP via WRDS for share prices and shares outstanding. None of the 99 assets I examine delist over the sample, and each asset becomes eligible for investment by the allocators once there is pricing data for each day of an estimation window.

I run causal discovery on the joint panel of the 99
assets and 33 plausibly exogenous macro-financial series (``drivers'') from FRED and
Yahoo Finance, including Treasury rates and slopes, credit spreads, commodities, FX
pairs, volatility indices, and international-equity ETFs. 
This means that the asset-to-asset submatrix is estimated conditional on the macro drivers, which removes the co-movement linearly explained by the drivers. I exclude series that are clearly non-exogenous, such as sector ETFs, US index returns and factor portfolios.

SVARs require weakly stationary inputs, so I transform time series accordingly: log-returns for price-like series, first differences for yields and spreads, year-on-year changes for monthly macro levels (which also removes seasonality), and no transform for the volatility indices, which are already stationary. 
The same transformations are applied in every estimation window to ensure a consistent definition for the variables.
I then check stationarity with ADF and KPSS tests. 

\subsection{Causal discovery}
\label{sec:discovery-run}

I use DYNOTEARS as my primary causal discovery method because it meets three key requirements.
First, the method must be scalable to process the ${\sim}132$-variable joint panel in reasonable time, across every monthly rebalance.
Second, it must be able to accept the ``directional prior'' ex ante, rather than removing asset-to-driver edges post hoc. This directional prior is that macro drivers can cause assets, but not vice versa, as the drivers are supposed to be exogenous. Thus, there are no edges from assets to drivers.
Third, the method must return a DAG. The allocators I use (see Section~\ref{chap:allocators}) propagate shocks along every directed path in the graph. If the graph is acyclic, I can sum the results exactly (Section~\ref{sec:sem}). 
A DAG also has a topological order, which I use in place of a clustering dendrogram in the ordering-based allocators I develop.
As Section~\ref{sec:causal-discovery} noted, VARLiNGAM's main weakness is that, at longer estimation windows, its identifying assumption of heavy-tailed residuals weakens. Ridge-Granger likewise has a weakness in that it does not guarantee an acyclic graph, and its coefficients are not structural effect sizes (Section~\ref{sec:comparators}).

At each rebalance date $t$, I fit DYNOTEARS on
a joint panel of drivers and assets $X=[\,D \mid A\,]$, with the rows representing the trading days in the estimation window
(there are four estimation window sizes, from 189 to 504 trading days; see Section~\ref{sec:measurement}), with a one-day lag.
As mentioned in Section~\ref{sec:causal-investing}, lags over 1 are negligible at these horizons.
I standardise all the variables in each window because, as mentioned in Section~\ref{sec:causal-discovery}, NOTEARS methods are scale dependent. I apply the same standardisation for the fits across the other two comparator discovery methods too, and convert the graph back to return units in Section~\ref{sec:sem}.
This fitting yields two matrices: a contemporaneous matrix $W$ and a lagged matrix $A_1$, both over the joint variable set.
Throughout the paper, $M[i,j]$ defines the effect of variable $i$ on variable $j$.
I list the hyperparameters in Appendix~\ref{app:hyper}.

I enforce the directional prior by populating DYNOTEARS' list of forbidden edges with all asset-to-driver links at each lag, forcing them to zero.
For VARLiNGAM I enforce the prior in both blocks. I fit the lagged block myself as a ridge VAR where each driver only regresses on lagged drivers, so no asset-to-driver lag coefficient can be non-zero. I pass the same restriction to DirectLiNGAM as a prior-knowledge matrix when it identifies the contemporaneous block $B_0$. I run VARLiNGAM without its own pruning step, and as a safeguard I zero any asset-to-driver mass that remains after the fit.
Refitting DYNOTEARS without the prior on six windows spanning 2008 to 2022 puts 36--48\% of the driver--asset edge mass on implausible asset-to-driver edges, so the prior binds.

\subsection{The asset graph}
\label{sec:asset-graph}

The ablation design requires each allocator to receive the same graph.
Therefore, at each rebalance, I take the asset-to-asset submatrix from the fitted contemporaneous matrix, and record whether it is acyclic. I also prune any edges with weights below a certain threshold: by default, the threshold is $0$ (no edges are pruned), but I sweep through multiple values in Section~\ref{sec:robustness}. 
To obtain the covariance from the structural equation model (SEM) (Section~\ref{sec:sem}), I calculate structural residual variances for each asset over the estimation window.

I find that the
DYNOTEARS graphs have an average of ${\sim}614$--$673$ directed edges across the 4 estimation windows, representing a
density of $6.3$--$6.9\%$ out of the possible ordered pairs of assets.
They have a mean
longest directed path of approximately $50$ assets (out of a total of 99), indicating quite deep chains of causal influence.
I find that
VARLiNGAM's unpruned contemporaneous matrix, $B_0$, has a much higher density of $32\%$ at the 189-day window; this seems to show a saturation effect with a lot of spurious or uninformative edges.

Because I lack a ground-truth causal structure, I evaluate the credibility of the discovered DAGs holistically: examining whether the directional prior has a quantifiable effect, how robust and non-degenerate the graphs are, and to what extent the different discovery methods disagree (see Section~\ref{sec:comparators}). I also indirectly evaluate the graphs based on the ablation testing that relies on them as inputs (see Sections~\ref{chap:allocators} and~\ref{chap:evaluation}).

\subsection{Alternative graph discovery algorithms}
\label{sec:comparators}

To check whether my results are robust, I use two alternative graph discovery algorithms to provide different classes of graphs to the allocators.

VARLiNGAM \cite{hyvarinen2010varlingam} identifies the instantaneous matrix $B_0$ through non-Gaussian residuals, rather than through a score, as discussed in Section~\ref{sec:causal-discovery}. The asset-only submatrix of $B_0$ is a DAG by construction.
Owing to its entirely different discovery mechanism, the resulting graphs are very different from those generated by DYNOTEARS: the differences in density are discussed in Section~\ref{sec:asset-graph}.

I also use a second comparator, though only as a narrow control at the 252-day estimation window on a small set of allocators. I call it ``ridge-Granger'' as it is a VAR(1) regression (used for uncovering Granger causality) with a ridge-regularisation term. 
To generate a graph from this regression, I define the edge weight from asset $i$ to asset $j$ as the absolute value of the lagged coefficient of $i$ in $j$'s regression. To allow for comparison between ridge-Granger and DYNOTEARS, I drop the lowest-weight edges from the ridge-Granger graph to ensure the graphs have equal density.
Note that comparing ridge-Granger to DYNOTEARS allows for the separation of directional information from edge magnitudes: its directions are informative because assets' movements can be predicted, but its edge weights are not structural effect sizes because they are just lagged predictive coefficients that have been ridge-regularised.
Because ridge-Granger does not necessarily produce an acyclic graph, total effects have to be approximated, and cycles have to be removed before the matrices are used by the ordering-based allocators.

These two discovery mechanisms therefore provide a robustness check: 
VARLiNGAM would not replicate any performance effect that relies on DYNOTEARS' particular edge directions, and ridge-Granger serves as a control to check whether directional information alone, rather than the precise edge magnitudes, is sufficient.
Section~\ref{sec:decomposition} tests this.

\section{Incorporating edge directions into hierarchical allocators}
\label{chap:allocators}

This section addresses the second challenge discussed in Section~\ref{sec:challenges}, incorporating edge directions into hierarchical allocators, and the design half of the first, attribution: Section~\ref{sec:ablation-design} sets out the ablation design that Section~\ref{chap:evaluation} executes.
I lay out several different allocators: two can only use the skeleton of the
graph, while the other three can also use its edge orientations.
The ablation compares these two groups of allocators, so that I can identify the performance effect attributable to edge orientation.
These new allocators also meet the requirement set out in
Section~\ref{sec:question}: that they can incorporate orientations. I discuss the problem in Section~\ref{sec:symm}, derive the relevant matrices and covariance in Section~\ref{sec:sem}, define the allocators in Sections~\ref{sec:baselines}--\ref{sec:orientation-family}, and finally discuss the ablation design in Section~\ref{sec:ablation-design}.

\subsection{Providing an asymmetric matrix to hierarchical allocators}
\label{sec:symm}

As discussed in Section~\ref{sec:challenges}, hierarchical allocation requires a
symmetric, non-negative distance matrix. This is problematic, because causal graphs are asymmetric by design, and simply symmetrising the adjacency matrix would remove the orientation information that is required to answer my research question.
I therefore adjust the allocation steps themselves so that they can take in the asymmetric matrix directly, while keeping everything else about the allocator fixed as far as possible to ensure that any performance differences can be attributed to this single change.
However, when the sample covariance matrix is replaced by $\Sigma_{\mathrm{SEM}}$, the covariance implied by the DAG, several things are changed beyond simply the orientation information; these other changes are discussed in Section~\ref{sec:orientation-family} and controlled for in Section~\ref{sec:orientation-role}.
There are two components of HRP where the asymmetric matrix can enter
directly. First, the dendrogram is just used to produce an asset ordering for the allocation step; using a DAG sidesteps that because it has a natural ordering along its causal chain. 
Second, HRP's bisection step uses variances from the covariance matrix; DAGs, when cast as a structural equation model, imply a covariance themselves anyway.

\subsection{Structural equation model}
\label{sec:sem}

The asset-only parts of the contemporaneous matrices produced by DYNOTEARS and VARLiNGAM,
$W$ and $B_0$, respectively, define a linear structural
equation model (SEM) \cite{wright1934path,bollen1989sem}
\begin{equation}   (I - M^\top)\, x = \varepsilon
\qquad\Longrightarrow\qquad   x = (I - M^\top)^{-1} \varepsilon, \label{eq:sem} \end{equation}
where $x$ is the vector of standardised asset returns, $\varepsilon$ is the vector of
structural shocks, and $(I - M^\top)^{-1}$ is the total-effect matrix, which I henceforth denote $T$ for brevity.
As the asset block is a DAG, its adjacency matrix is
nilpotent, which implies that the total-effect matrix $T = \sum_{k\ge 0}
(M^\top)^k$ has an exact value with at most $N$ terms.
Each entry $T_{ij} = \partial
x_i / \partial \varepsilon_j$ 
captures the total effect on asset $i$ of a unit shock to asset $j$ along all directed paths. 
The asymmetry of this matrix allows for orientation information to be used by the hierarchical allocators. The ridge-Granger control produces a graph that is not necessarily acyclic, so $T$ (and the corresponding SEM) has to be approximated by truncating the series after 10 terms (rescaling $M$ if its spectral radius is 1 or more).

The SEM also implies a covariance for the asset returns. Since
$x = T\varepsilon$, the covariance of $x$ is $T\,\Sigma_\varepsilon\,T^\top$,
where $\Sigma_\varepsilon$ is the covariance of the structural shocks.
The SEM assumes that shocks are independent, so every off-diagonal
entry of $\Sigma_\varepsilon$ is zero. I therefore set
$\Sigma_\varepsilon = \operatorname{diag}(\sigma^2_{\varepsilon,i})$.
I do not estimate the off-diagonal entries, because that would involve measuring correlations between residuals, thereby feeding the correlation matrix used by \corrhrp{} back into the graph's covariance, which would not test the graph's structure alone.
I convert the above into return units in order to get the assets' de-standardised cluster variances (which are necessary for the recursive bisection step of allocation). This gives
\begin{equation}
\Sigma_{\mathrm{SEM}}   \;=\; D_\sigma \big( T\, \Sigma_\varepsilon\,
T^\top \big) D_\sigma ,   \label{eq:sigma-sem} \end{equation} 
where $D_\sigma$ converts from the standardised values that the graph was fitted in to the correct return units. 
The variances of the shocks are estimated from the residuals, $E = \tilde{X} (I - M)$, where $\tilde{X}$ is the standardised panel of asset returns in the estimation window.
Note that $\Sigma_{\mathrm{SEM}}$ defines the covariance of the assets implied by the structural model (the $\Sigma(\theta)$ of SEM, or the reduced
form in SVAR terminology), while $\Sigma_\varepsilon$ above defines the covariance of the structural shocks.
There are two caveats to the above. Firstly, the SEM over the assets marginalises over the drivers (and lagged observations), which means that $\varepsilon$ absorbs driver-explained variance. As a result, $\Sigma_\varepsilon$ is an approximation.
Also, to avoid ill-conditioning, I add a small value to every diagonal entry of $\Sigma_{\mathrm{SEM}}$.

\subsection{Correlation and skeleton baselines}
\label{sec:baselines}

The correlation baseline, \corrhrp{}, is standard HRP \cite{lopezdeprado2016hrp}.
The clustering distance $\sqrt{\tfrac{1}{2}(1-\rho_{ij})}$ is calculated from the estimation window's sample covariance, which is itself used for bisection. 
Note that standard HRP does not use a causal graph anywhere. All of the relative performance effects of other allocators that I report here are measured relative to this \corrhrp{} baseline.

The skeleton allocator, \skelhrp{}, replaces the correlation matrix with the graph, through a distance that is transpose-invariant.
The clustering distance between assets is $D_{ij} =
\lVert e_i - e_j \rVert_2$, where $e_i$ is an asset's edge profile $[\,M_{i\cdot},\, M_{\cdot i}\,]$. This is provably invariant to reversing all edges simultaneously.
However, since $e_i$ distinguishes between outgoing and incoming edges, reversing a single edge's direction can move mass between the two parts of the vector, thereby changing the distances.
\skelhrp{} therefore is aware of the incoming and outgoing edges of each asset, which is information the undirected skeleton alone should not contain. 
I therefore define a robustness check with a stricter guarantee, \undirhrp{}, below, but keep \skelhrp{} as the primary skeleton-only reference, for two reasons.
First, the treatments of Section~\ref{sec:orientation-family} are designed to differ from the control in a single aspect each. Concretely, \semcovhrp{} and the
mechanism controls in Section~\ref{sec:orientation-role} keep the dendrogram from \skelhrp{} and change only the covariance.
Using \undirhrp{} as the skeleton-only baseline would also change the dendrogram, so would not offer an identifiable effect. Second, \skelhrp{} aligns with the allocator I used in the first phase of my research (see Appendix~\ref{app:hsp}).

\undirhrp{} uses a different symmetrisation approach.
It uses the same pipeline, but the clustering distance is built from
$\tfrac{1}{2}(|M|+|M^\top|)$, the undirected weighted adjacency: the
off-diagonal entries are divided by their maximum so that they lie in
$[0,1]$, and the distance is one minus that scaled similarity.
This defines distance between assets based on the strength of the edge joining them directly, compared to \skelhrp{}, which defines two assets to be close when their $e_i$ vectors are similar (i.e. if they share incoming and outgoing edges, even if no edge joins them directly).
\undirhrp{} carries no orientation information
at all, because undirected adjacency is not changed when any edge is reversed.
Rather than acting as a second primary control, it acts as a robustness check on the skeleton control: Section~\ref{sec:decomposition} repeats the skeleton contrast against it to see whether the performance of the skeleton-only allocators depends on the definition of the ``skeleton'' itself.

\subsection{Edge-orientation-aware allocators}
\label{sec:orientation-family}

As discussed in Section~\ref{sec:symm}, HRP has two places where information from causal graphs can be substituted in.
The clustering step only exists to create an ordering of the assets, while the allocation step uses that ordering (splitting the ordered list by position), as well as a covariance.
Therefore the causal graph can enter into HRP either at the ordering or covariance: see Table~\ref{tab:crossing} for the different options.

\begin{table}[htbp]
  \centering
  \caption{Where the causal graph can enter HRP, and the resulting allocators.}
  \label{tab:crossing}
  \setlength{\aboverulesep}{0pt}
  \setlength{\belowrulesep}{0pt}
  \setlength{\extrarowheight}{2.5pt}
  \begin{tabular}{l | l l}
    \toprule
    \diagbox[height=2.4\line]{Ordering}{Covariance}
      & sample covariance & $\Sigma_{\mathrm{SEM}}$ \\
    \midrule
    clustering on skeleton & \skelhrp{} (control) & \semcovhrp{} \\
    topological sort           & \topohrp{}           & \toposemcovhrp{} \\
    \bottomrule
  \end{tabular}
\end{table}

The rows indicate what produces the ordering. The first retains HRP's clustering step, but uses a distance derived from the graph's skeleton. The second row eliminates the clustering step entirely. This is because a DAG is inherently ordered, and therefore does not require a dendrogram to produce an ordering.
Each of the allocators differs from the control in incremental aspects: the covariance alone for \semcovhrp{}, the ordering alone for \topohrp{}, and both for \toposemcovhrp{}.
These differences mean that I can identify whether any of them has an impact on performance. All three allocators are long-only, fully invested, and deterministic, and they all take the same inputs as the controls: the graph $M$ and the returns estimation window. They are also all novel, to my knowledge.

\semcovhrp{} uses the covariance implied by the DAG's SEM form for recursive bisection at the allocation step, and keeps its clustering the same as in \skelhrp{}.
Taking the difference between the two allocators thereby isolates the effect of swapping the sample covariance for $\Sigma_{\mathrm{SEM}}$ at the allocation step.
Orientation information enters into this allocator due to the fact that shocks can asymmetrically propagate through $T$ (see Eq.~\eqref{eq:sigma-sem}). However, $\Sigma_{\mathrm{SEM}}$ has other properties, such as the imposition of
independent structural shocks (which removes the common market factor between asset series), and the fact that off-diagonal entries are shrunk.
Therefore the difference between the two allocators cannot cleanly attribute a performance impact to edge directions alone: the de-factored and Ledoit--Wolf
controls in Section~\ref{sec:orientation-role} isolate direction from the two properties above.

\topohrp{} and \toposemcovhrp{} remove HRP's clustering step, as DAGs can directly provide the ordering required as an input to bisection.
I sort the assets from upstream to downstream using a topological sort of $M$. 
Where ties exist, I order the tied assets by total downstream influence
$\sum_r |T_{ri}|$ (and then by name). 
The standard HRP recursive bisection runs using the resulting ordering, with the covariance being either the sample, in the case of \topohrp{}, or $\Sigma_{\mathrm{SEM}}$, in the case of \toposemcovhrp{}.
This sidesteps the problem of being unable to cluster on asymmetric matrices.

\subsection{Ablation design}
\label{sec:ablation-design}

The ablation design requires the same graph $M$ to be provided to all the allocators in Table~\ref{tab:variants}, alongside other basics like the set of assets, the transaction costs, and the estimation window.
Thus, the only factor that changes is
how the graph enters the allocator.
The resulting decomposition consists of two differences: $\skelhrp{}-\corrhrp{}$ (the marginal impact of the skeleton over HRP) and $\oriented{}-\skelhrp{}$ (the marginal impact of the edge orientations over the skeleton allocator, where \oriented{} denotes any of the 3 edge-orientation-aware allocators I consider). Adding the two yields the total performance impact of using causal graphs (both their edge orientations and their skeleton) over HRP.

Table~\ref{tab:variants} lists the primary family: the baseline \corrhrp{}, plus the 5
graph-consuming allocators that form the narrowed family of Section~\ref{chap:evaluation}, namely the skeleton control \skelhrp{}, its robustness check \undirhrp{}, and the three edge-orientation-aware allocators. 
Each of these is run under both discovery methods (DYNOTEARS and VARLiNGAM), at each of
the four estimation windows (189, 252, 378 and 504 days). This yields the full set of results in Section~\ref{chap:evaluation}.
Separately, I also use the ridge-Granger control, and a robustness check across sparsity and cost parameters.
In Appendix~\ref{app:tables}, I also include a run across a wider family of 7 graph-consuming allocators (which was actually conducted before narrowing to the main 5-allocator family), and a second family of hierarchical allocators (HERC). I also added mechanism controls (two direction-free covariance controls and a graph-free partial-correlation skeleton control) during the evaluation, to further isolate the drivers of the performance effects behind the edge-orientation-aware and skeleton-only allocators. See Table~\ref{tab:allocator-map} in Appendix~\ref{app:tables} for full details. 
Further details and reasoning behind the narrowing and the additional controls
are discussed in Section~\ref{sec:caveats}. To avoid the multiple comparisons problem, every evaluated configuration is included in the \rsNtrials\ number that the Deflated Sharpe
ratio corrects for. 

\begin{table}[t]
  \centering
  \caption{The primary allocator family.}
  \label{tab:variants}
  \small
  \setlength{\tabcolsep}{4.5pt}
  \begin{tabular}{l l l l}
    \toprule
    Allocator & Ordering from & Covariance for bisection & Reads orientation? \\
    \midrule
    \corrhrp{} & correlation distance & sample & no (graph-blind) \\
    \skelhrp{}            & graph embedding      & sample & no (transpose-inv.) \\
    \undirhrp{}           & $\tfrac{1}{2}(|M|{+}|M^\top|)$ & sample & no (per-edge inv.) \\
    \semcovhrp{}            & graph embedding      & $\Sigma_{\mathrm{SEM}}$ & allocation step \\
    \topohrp{}            & topological order    & sample & ordering step \\
    \toposemcovhrp{}           & topological order    & $\Sigma_{\mathrm{SEM}}$ & both steps \\
    \bottomrule
  \end{tabular}
  \floatnote{For all other allocators, see Table~\ref{tab:allocator-map}, Appendix~\ref{app:tables}.}
\end{table}

\section{Decomposing the effect}
\label{chap:evaluation}

In this section, I address the first and third challenges discussed in Section~\ref{sec:challenges}.
In Section~\ref{sec:measurement}, I measure the performance effect.
Section~\ref{sec:graph-vs-correlation} answers the first part of the research question: whether the causal
graph improves on the correlation matrix at all.
Section~\ref{sec:decomposition} then answers the second part by running the ablation design laid out in Section~\ref{sec:ablation-design}, decomposing the effect into the components attributable to the graph's skeleton and its edge orientations. It also runs the mechanism controls behind both components and assesses how robust the decomposition is to different parameters, discovery methods, and allocators.
In Sections~\ref{sec:multiplicity} to \ref{sec:caveats}, I conduct statistical inference in the face of small point estimates and large standard errors (owing to the noise in the 18-year sample). In Section~\ref{sec:multiplicity}, I account for the fact that I search over \rsNtrials\ assessed configurations to avoid the multiple comparisons problem.
In Section~\ref{sec:oos} I see how the allocators perform out-of-sample, and finally Section~\ref{sec:caveats} sets out the threats to validity.

\subsection{Measurement procedure}
\label{sec:measurement}

I use the same backtest to measure the performance of every allocator configuration I consider. I use a fixed 99-asset subset of the S\&P~500, monthly rebalancing over 2007--2024, estimation windows of 189, 252, 378 and 504 trading days (both for fitting the causal graphs and for estimating the sample covariance), and impose transaction costs of 5\,bps per unit of one-way turnover (I also check other transaction cost values for robustness in Section~\ref{sec:robustness}).
From January 2007 to November 2024, inclusive, there are 215 monthly rebalances at each estimation window. Because there is a lag before the longer estimation windows can be used for rebalancing portfolios, the 378- and 504-day windows have only 209 and 203 rebalances that are fed by the DAGs, respectively; the earliest months just use equal weighting.
I define ``performance'' to be out-of-sample, net-of-cost, annualised Sharpe ratio of daily portfolio returns: see Table~\ref{tab:return-terms} in Appendix~\ref{app:tables} which reports this annualised Sharpe alongside compound annual growth rate (CAGR), maximum drawdown and mean turnover.

I perform inference using a variety of statistics.
For pairwise contrasts, I use the Politis--Romano stationary block bootstrap \cite{politis1994stationary} with 10{,}000 resamples and a mean block length of 21 days, resampling the joint return panel so as to maintain cross-strategy correlation.
The two-sided $p$-value I report is the percentile one: one minus the highest percentage confidence interval that still excludes zero.
I find that daily returns are highly non-normal, with an excess kurtosis of
${\approx}\,\rsKurtosis$.
I assessed \rsNtrials\ configurations in total. To correct for the potential multiple comparisons problem, I apply several tests discussed in Section~\ref{sec:multiplicity}: the Probabilistic and Deflated Sharpe
ratios \cite{bailey2012psr,bailey2014deflated}, Hansen's SPA \cite{hansen2005spa} (for every comparison of a family of allocators against a baseline), and the $90\%$ Model Confidence Set \cite{hansen2011mcs}.
Note that the latter two tests use the daily net return as their performance metric (the loss is its negative), so their hypotheses concern mean return rather than the Sharpe ratio.

\subsection{Causal graphs' performance compared to correlation}
\label{sec:graph-vs-correlation}

The research question's first part asks whether replacing the correlation matrix with a causal graph improves hierarchical allocation at all.
To answer this, I test for a difference between the \corrhrp{} baseline and the causal graph allocators.
\corrhrp{}'s Sharpe ratios are $\rsCorrSharpeWone$,
$\rsCorrSharpe$, $\rsCorrSharpeWthree$ and $\rsCorrSharpeWfive$ for the
189-, 252-, 378- and 504-day windows, respectively.
At these respective windows, \semcovhrp{} (the edge-orientation-aware allocator that uses the implied covariance of the DAG at the allocation step) differs from that baseline by $-0.004$, $-0.017$, $+0.006$ and $+0.013$ ($p=0.36$--$0.82$, see Table~\ref{tab:decomp}).
Similarly, \toposemcovhrp{}, which also uses the DAG's topological order at the ordering step, differs from \corrhrp{} by $-0.003$, $-0.021$, $+0.012$ and $+0.016$, respectively ($p=0.25$--$0.87$).
There are a total of 20 comparisons between the DYNOTEARS-fed
allocators and \corrhrp{} (5 allocators at 4 windows). Only 6 of these DYNOTEARS comparisons are positive, and all of these 6 are at the two longer windows, and only 1 of the 20 VARLiNGAM comparisons is positive.
At the two shorter windows, \corrhrp{} outperforms every graph-fed allocator, regardless of the discovery method.

In summary, the causal graph does not improve on the correlation matrix, and at the shorter windows it is worse. Within the DYNOTEARS primary family no shortfall reaches significance (the closest is \skelhrp{} against \corrhrp{} at 252 days, $-0.028$ at $p=0.056$); under VARLiNGAM graphs, \semcovhrp{} falls short of the baseline by $-0.071$ at 189 days ($p=0.036$) and $-0.042$ at 252 days ($p=0.046$). The answer to the first part of the research question is therefore no.
I conduct a deeper check of this by running Hansen's SPA test (where the
null hypothesis is that no allocator in the family outperforms
\corrhrp{}) on two families at each window, with the candidates
drawn from both discovery methods.
The test does not reject the null at any window, for
either the full family of 7 allocators or the narrowed family of 5, nor
when the 4 windows are pooled: see Section~\ref{sec:multiplicity} for further details.

Though the total effect of the causal graph is zero, the decomposition in Section~\ref{sec:decomposition} identifies the relative contributions of the two separate aspects of the graph.

\subsection{Decomposing performance into skeleton and edge orientations}
\label{sec:decomposition}

\subsubsection{Decomposition across estimation windows}

\begin{figure}[tbp]
  \centering
  \includegraphics[width=0.99\textwidth]{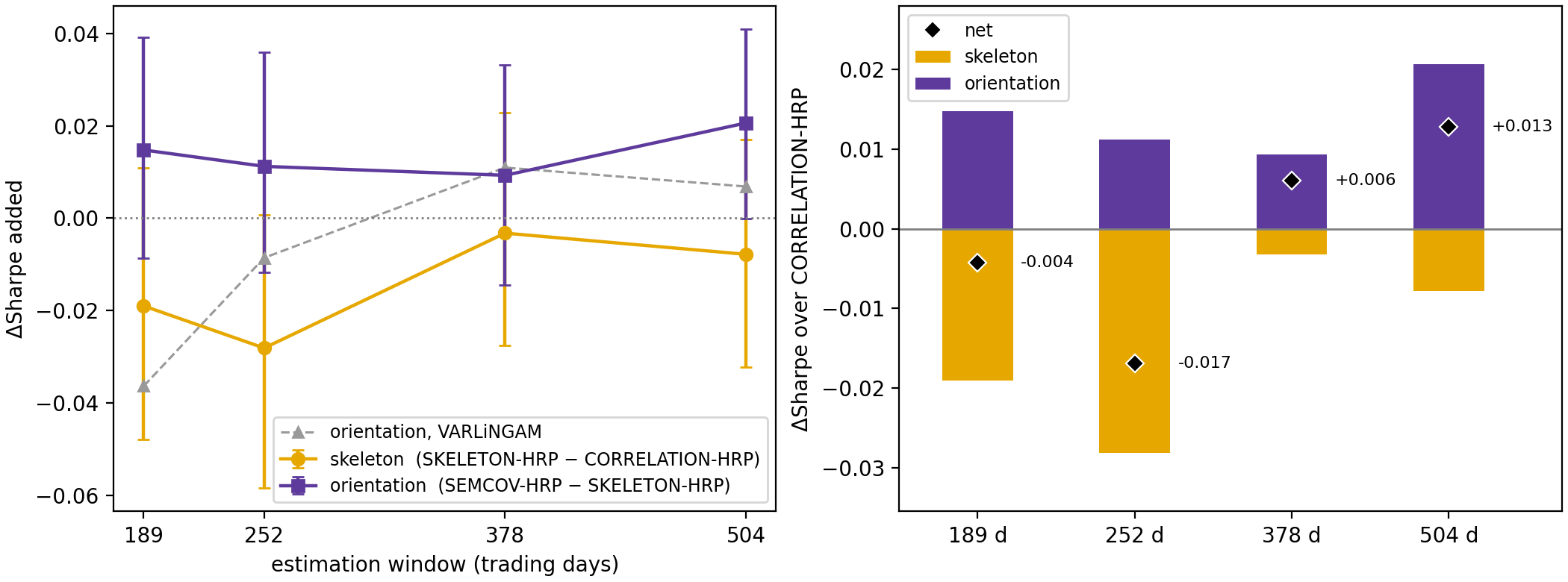}
  \caption{The decomposed performance effects of two components of the DAG (DYNOTEARS).}
  \label{fig:gradient}
  \floatnote{Left: the skeleton's contribution (\skelhrp{} $-$ \corrhrp{}) and the orientations' contribution (\semcovhrp{} $-$ \skelhrp{}), with $95\%$ block-bootstrap confidence intervals. The VARLiNGAM orientation series displays the same contrast (\semcovhrp{} $-$ \skelhrp{}).
  Right: the same two contributions stacked together, and the net effect (\semcovhrp{} $-$ \corrhrp{}).}
\end{figure}

\begin{figure}[tbp]
  \centering
  \includegraphics[width=0.99\textwidth]{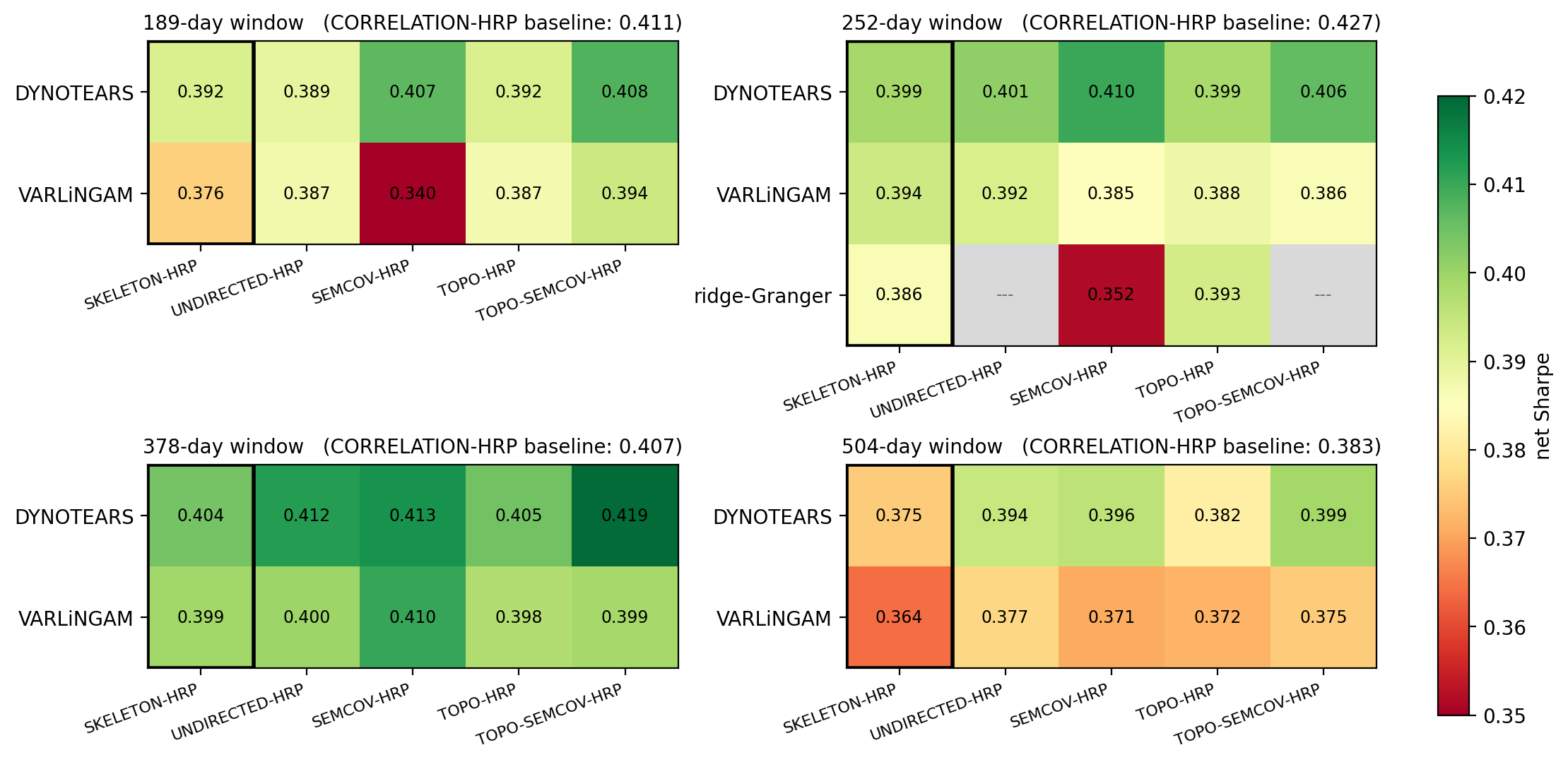}
  \caption{Net Sharpe by allocator and causal discovery method at the 4 estimation windows.}
  \label{fig:heatmap}
  \floatnote{The \corrhrp{} net Sharpe values are in the title of each window. Ridge-Granger is used as an ancillary discovery method, for specific allocators at the 252-day window only.}
\end{figure}

\begin{figure}[tbp]
  \centering
  \includegraphics[width=0.99\textwidth]{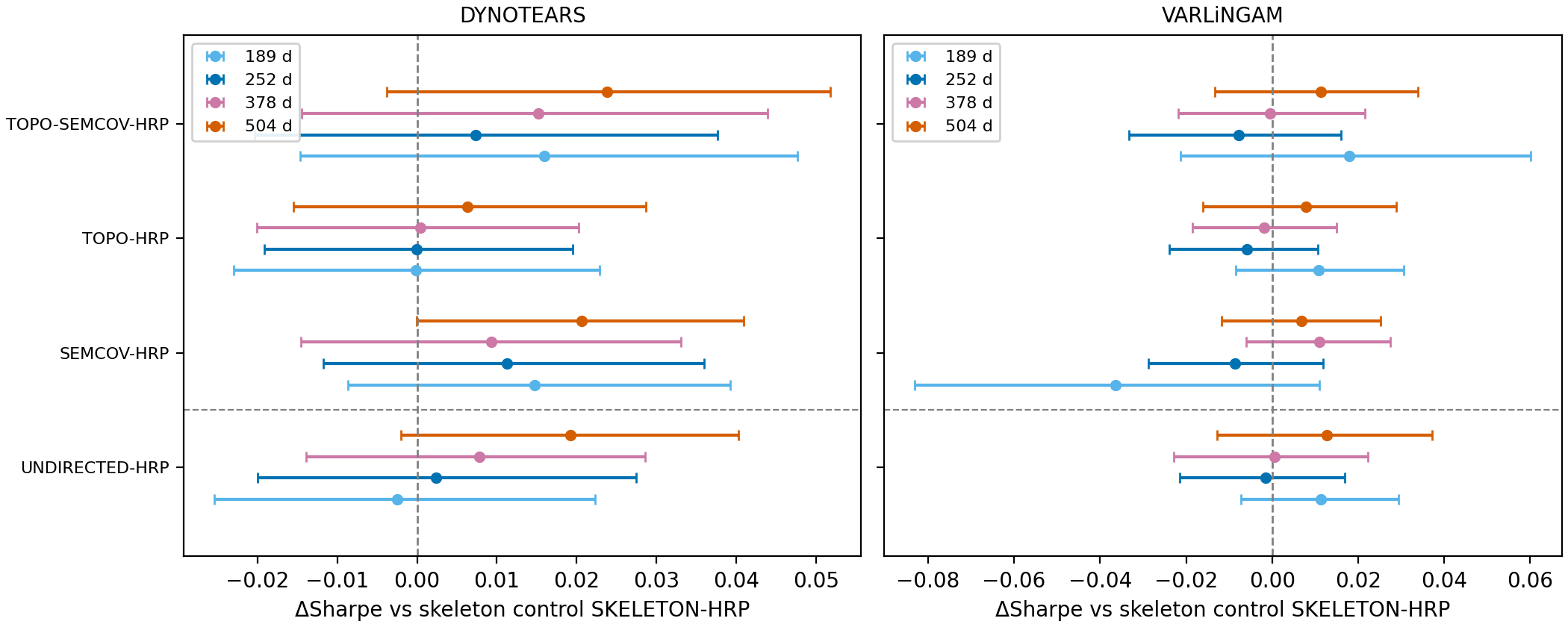}
  \caption{$\Delta$Sharpe of each edge-orientation-aware allocator and \undirhrp{} against \skelhrp{} at all 4 windows.}
  \label{fig:forest}
  \floatnote{$95\%$ block-bootstrap confidence intervals. \undirhrp{} is a stricter skeleton-only allocator than \skelhrp{} as it carries no orientation information, but it is not in keeping with the incremental nature of the ablation design.}
\end{figure}

The paper's central result is displayed in Figure~\ref{fig:gradient}, which shows the relative performance effects of the two components of the decomposition across the 4 estimation windows.
The numerical values are in Table~\ref{tab:decomp}, and Figure~\ref{fig:heatmap} gives the
full matrix across allocators, discovery methods, and estimation windows.
The second part of the research question, how much of any improvement is attributable
to the graph's skeleton versus its edge orientations, is answered by two features.
First, the skeleton's marginal effect (\skelhrp{} minus \corrhrp{}) is negative at every window: $-0.019$, $-0.028$, $-0.003$, and $-0.008$ at the 189-, 252-, 378-, and 504-day windows, respectively. 
Second, the edge orientations' marginal effect (\semcovhrp{} minus \skelhrp{}) is
positive at every window: $+0.015$, $+0.011$, $+0.009$, and $+0.021$, respectively. 
All 12 of the comparisons between DYNOTEARS edge-orientation-aware allocators and \skelhrp{} are non-negative to three decimal places (Figure~\ref{fig:forest}).
Though the point estimates of the performance effects for the two components above have opposite signs, none of them is statistically distinguishable from zero (Table~\ref{tab:decomp}) at any window.
I also conduct Hansen's SPA test to compare the family of edge-orientation-aware allocators to \skelhrp{} under both causal discovery methods. I find that it does not reject the null that none of them outperforms \skelhrp{} at any window; see Section~\ref{sec:multiplicity} for further details.

The skeleton-only allocator discussed above is \skelhrp{}: as mentioned in Section~\ref{sec:baselines}, the alternative symmetrised allocator is \undirhrp{}, which is more strongly orientation-blind. Using this allocator, the skeleton's contribution over \corrhrp{} becomes $-0.022$, $-0.026$, $+0.005$ and $+0.011$ across the 4 windows, and the relative contribution of the edge orientations over \undirhrp{} is $+0.017$, $+0.009$, $+0.001$ and $+0.001$ for \semcovhrp{} and $+0.018$, $+0.005$, $+0.007$ and $+0.005$ for \toposemcovhrp{}. Note that these are point estimates and not formally tested. 
At the two shorter windows, the signs of the results under both skeleton-only allocators are the same, but at the two longer windows they differ: \undirhrp{} outperforms
\skelhrp{} by $+0.008$ and $+0.019$ (though not significantly), which is why the orientation effect at the 378- and 504-day windows essentially disappears when compared to \undirhrp{}.
Thus, the overall finding is that edge orientations yield a small positive improvement relative to the skeleton at the shorter estimation windows: their relative performance at the longer windows holds only against \skelhrp{}.

\begin{table}[tbp]
  \centering
  \caption{The decomposed performance effects (pairwise $\Delta$Sharpe, DYNOTEARS).}
  \label{tab:decomp}
  \small
  \setlength{\tabcolsep}{5pt}
  \begin{tabular}{l c c c c}
    \toprule
    & \multicolumn{4}{c}{Window (days)} \\
    \cmidrule(lr){2-5}
    Contrast & 189 & 252 & 378 & 504 \\
    \midrule
    skeleton: \skelhrp{} $-$ \corrhrp{} & $-0.019$ & $-0.028$ & $-0.003$ & $-0.008$ \\
      & {\footnotesize(0.21)} & {\footnotesize(0.056)} & {\footnotesize(0.80)} & {\footnotesize(0.56)} \\
    \addlinespace[2pt]
    orientation (allocation step): \semcovhrp{} $-$ \skelhrp{} & $+0.015$ & $+0.011$ & $+0.009$ & $+0.021$ \\
      & {\footnotesize(0.22)} & {\footnotesize(0.35)} & {\footnotesize(0.45)} & {\footnotesize(0.051)} \\
    \addlinespace[2pt]
    orientation (both steps): \toposemcovhrp{} $-$ \skelhrp{} & $+0.016$ & $+0.007$ & $+0.015$ & $+0.024$ \\
      & {\footnotesize(0.31)} & {\footnotesize(0.61)} & {\footnotesize(0.33)} & {\footnotesize(0.086)} \\
    \midrule
    total: \semcovhrp{} $-$ \corrhrp{} & $-0.004$ & $-0.017$ & $+0.006$ & $+0.013$ \\
      & {\footnotesize(0.82)} & {\footnotesize(0.36)} & {\footnotesize(0.73)} & {\footnotesize(0.44)} \\
    \addlinespace[2pt]
    total: \toposemcovhrp{} $-$ \corrhrp{} & $-0.003$ & $-0.021$ & $+0.012$ & $+0.016$ \\
      & {\footnotesize(0.87)} & {\footnotesize(0.25)} & {\footnotesize(0.50)} & {\footnotesize(0.33)} \\
    \bottomrule
  \end{tabular}
  \floatnote{Bootstrap $p$-values in parentheses. Total effect $=$ skeleton effect $+$ orientation effect.}
\end{table}

\FloatBarrier
\subsubsection{Drivers of the edge orientations' performance}
\label{sec:orientation-role}

\semcovhrp{} and \toposemcovhrp{} perform fairly evenly across estimation windows (with Sharpe ratios of $0.396$ to $0.413$ and $0.399$ to $0.419$, respectively). Figure~\ref{fig:gradient} shows that the orientations' marginal contribution (relative to \skelhrp{}) is positive at every window. 
This indicates that edge orientations partly counteract the graph skeleton's negative effect relative to the correlation baseline, but do not yield a performance gain in absolute terms.
The edge orientations enter \semcovhrp{} only through the DAG's implied (SEM) covariance $\Sigma_{\mathrm{SEM}}$. This has two properties that are unrelated to the edge directions themselves but could nevertheless be driving the performance gain.
First, the implied covariance is a structured estimator that has fewer parameters
than the sample covariance: in this way, it acts like a shrinkage estimator.
Second, it assumes that shocks are independent: this removes the common market factor.
Neither of these properties necessarily requires a graph. I therefore test whether either of them is responsible for the performance gain by using two controls which replace \skelhrp{}'s sample covariance at the allocation step but do not use edge orientations. 
The first control, \lwhrp{} (see Table~\ref{tab:allocator-map}), applies Ledoit--Wolf shrinkage \cite{ledoitwolf2004} to the covariance matrix to see whether shrinkage drives the performance effect. 
The second control, \dfhrp{}, uses a de-factored covariance for allocation, namely, the covariance of the residual from a regression of each asset's return on the average return across all 99 assets within the window. This regression removes the common market factor.

I find that shrinkage does not reproduce the gain: the Ledoit--Wolf
control has a lower performance than \skelhrp{} in every window ($\rsDlwSharpeWone$,
$\rsDlwSharpe$, $\rsDlwSharpeWthree$, $\rsDlwSharpeWfive$ across the 4
windows), and is outperformed by \semcovhrp{} (by between $+0.016$ and $+0.040$).
However, the de-factoring control does reproduce the edge-orientation-aware allocator's performance gain. In all 4 windows, its performance
($\rsDdfSharpeWone$, $\rsDdfSharpe$, $\rsDdfSharpeWthree$,
$\rsDdfSharpeWfive$) matches that of \semcovhrp{}, with the difference between the two being between $-0.011$ and $+0.001$.
This indicates that the performance gain from using the implied, rather than sample,
covariance is due to its assumption of independent shocks,
which eliminates the common market factor, rather than to the edge
directions themselves.
However, reliable edge magnitudes are still necessary for the SEM. As discussed in the robustness checks below, edge-orientation-aware allocators that use VARLiNGAM graphs, whose edges are less reliable, show no systematic gain over the skeleton. The ridge-Granger graphs, whose edge magnitudes are incorrect, do not yield any consistent effect: \topohrp{} ($0.393$) exceeds \skelhrp{} ($0.386$) while \semcovhrp{} ($0.352$) scores below it.

\subsubsection{Drivers of the skeleton's performance}
\label{sec:skeleton-role}

Next, I examine whether the negative performance effect (relative to \corrhrp{}) from using the graph's skeleton is specifically a property of the discovered causal graph. This need not be the case: in theory, any other conditional-dependence structure of similar sparsity could yield the same effect, with no need for causal discovery.
I test this using \pcorrhrp{} (Table~\ref{tab:allocator-map}) as a control, which replaces the discovered skeleton in \skelhrp{} with a pruned partial-correlation
matrix with the same number of non-zero entries. 
To do this, I first obtain a 
precision matrix by inverting a Ledoit--Wolf covariance (which shrinks sample covariances towards an identity matrix), remove the smallest-magnitude entries, and then apply a threshold so that the number of non-zero entries matches the number of edges in the DAG.
Note that, because the partial-correlation matrix is symmetric, each asset pair occupies
two entries, whereas a DAG has one entry (edge) per pair. As a result, \pcorrhrp{} links half as many asset pairs as the DAG's skeleton, so it is possible that its sparsity is a driver of the difference reported below.
I find that \pcorrhrp{} matches the performance of \skelhrp{}, the skeleton-only allocator, at all 4 windows. At 252 days, its Sharpe is $\rsDpcSharpe$,
compared to \skelhrp{}'s $0.399$: the difference is insignificant ($+0.012$, $p=0.23$). 
At the other 3 windows, the
differences between the two are $-0.020$ ($p=0.07$), $-0.005$ and $+0.007$ ($p \geq
0.48$). 
Both of these skeletal allocators underperform the correlation baseline: 
the difference between \pcorrhrp{} and \corrhrp{} is $-0.039$ ($p=0.008$), $-0.016$, $-0.008$ and $-0.001$ across the 4
windows. 
Thus, I find that a partial-correlation network with the same number of entries as the DAG's skeleton replicates the performance of the latter, implying that the discovered graph itself adds nothing at the clustering step beyond what is provided by a generic sparse dependence network.

\subsubsection{Robustness checks}
\label{sec:robustness}

I perform several robustness checks to ensure that my results are consistent with different choices of parameters. The first is the sparsity threshold $\tau$: any edge weight that is below this threshold is pruned from the graph. I try a range of values, $\tau\in\{0,0.01,0.05,0.10\}$, and find net annualised Sharpe essentially unchanged. \topohrp{}, the allocator that uses the DAG's topological order at HRP's ordering step, sees its Sharpe go from $0.399$ to $0.402$ at the strictest threshold, while the skeleton control goes from $0.399$ to $0.395$, dipping to $0.386$ at $\tau=0.05$ (Table~\ref{tab:tau}).
The sweep did not cover \semcovhrp{}, the headline treatment, so its sensitivity to $\tau$ is untested, and I did not estimate the statistical significance of the differences above.

I also check my results against different transaction costs.
I find that the performance gain from the implied
covariance at the allocation step, \semcovhrp{}$-$\skelhrp{}, is robust to different transaction costs. At the 504-day window, the gain moves from $+0.020$ to $+0.023$ ($0.3864$ against $0.3638$) as
costs rise from 0 to 20\,bps (see Table~\ref{tab:cost}).
This is because \semcovhrp{}'s one-way turnover ($0.089$ per rebalance) is less than the control's ($0.107$).
By contrast, the gain at the ordering step erodes with cost at the 252-day window.
\toposemcovhrp{}'s turnover ($0.140$ at 252 days) is the second highest in the family, behind \undirhrp{}'s ($0.145$). As a result, its performance gain over \skelhrp{} at 252 days decreases from $+0.008$ at zero cost to $+0.005$ at 20\,bps.
Its gain at the 504-day window ($+0.024$ to $+0.022$) is robust to different transaction cost values.

I also use different causal discovery methods, as mentioned.
Using graphs generated by VARLiNGAM, the edge orientations give no systematic gain (6 of 12 comparisons are
positive, with a mean of $0.000$; see Figures~\ref{fig:forest} and \ref{fig:gradient}).
This shows that the performance decomposition is not agnostic to the causal discovery method.
VARLiNGAM's weakness is likely due to the fact that its unpruned contemporaneous matrix gets saturated with spurious edges, with a density of $32\%$ against DYNOTEARS' roughly $6$--$7\%$ (see Section~\ref{sec:asset-graph}), so the edge directions are not as informative.

I also use a ridge-Granger graph as a control on edge magnitudes.
DYNOTEARS graphs provide both the direction of each edge and its fitted magnitude, so the orientation gain cannot be attributed cleanly to one of the two. The ridge-Granger control preserves the ordering while degrading the magnitudes.
With its graphs, causal-ordered bisection is essentially unchanged (\topohrp{} $=0.393$ against DYNOTEARS' $0.399$), while \semcovhrp{} falls to $0.352$ (a gap of $-0.058$ against its DYNOTEARS value, $p=0.13$).
The orientation contribution therefore depends on reliably estimated edge magnitudes that can propagate shocks, not on the directions alone.

The final robustness check uses a different hierarchical allocator, HERC \cite{raffinot2018herc}, to test the proposition of Section~\ref{sec:symm} that any hierarchical allocator clustering on a symmetric matrix loses orientation information.
HERC shares HRP's clustering step and cluster variance measure, but splits assets by following the topology of the dendrogram rather than bisecting the leaf ordering, which tends to concentrate weights.
The three HERC cells mirror the decomposition's three HRP anchors: \corrherc{}, \skelherc{} and \semcovherc{} (Table~\ref{tab:allocator-map}).
The signs of both components generalise: under HERC the orientations add at every window and the skeleton costs at every window, as under HRP, but the sizes are larger and none of the contrasts is significant (Appendix~\ref{app:tables}). The size of the decomposition therefore depends on the allocator.
All 12 HERC configurations count towards \rsNtrials\ but are in neither SPA family.

\subsection{Multiplicity-aware inference}
\label{sec:multiplicity}

My statistical testing has to address the multiple-comparisons problem: the possibility that the best-performing configurations arise by chance out of the numerous trials I run.

The Probabilistic Sharpe ratios, which give the probability that each allocator's true Sharpe exceeds zero given the sample's skewness, kurtosis and length, are high (at 252 days, $\rsCorrPSR$ for \corrhrp{}, $\rsVprimePSR$ for \skelhrp{} and $\rsDonePSR$ for \semcovhrp{}), so non-normality does not threaten the sign of any level.
The Deflated Sharpe ratio accounts for selection bias: it compares each Sharpe ratio against the best that would be expected from \rsNtrials\ trials with a true Sharpe of zero, and reports the probability that the true Sharpe is above zero after that search; the conventional bar is $0.95$. No configuration reaches it. \corrhrp{} at the 252-day window is among the highest DSR values ($\rsCorrDSR$), and the spread between the best configuration and this baseline is under $0.02$, so the DSR speaks to the individual levels and does not rank the allocators.

To allow for comparisons over a family of allocators, I use Hansen's SPA test.
The null hypothesis is that no allocator in the family outperforms the baseline, and the test accounts for the fact that the best-performing member was chosen from the family.
I apply the test to two
families: first, the family of 7 graph-consuming allocators against the \corrhrp{} baseline, to test whether using the causal graph at all beats the baseline correlational allocator; second, just the 5 edge-orientation-aware allocators versus the DYNOTEARS skeleton-only allocator, \skelhrp{}, to test whether there is an additional gain from the graph's edge orientations above that of the skeleton (Section~\ref{sec:decomposition}).
Both families contain configurations that use both VARLiNGAM and DYNOTEARS as causal discovery algorithms, so the second family also compares the VARLiNGAM edge-orientation-aware allocators to the DYNOTEARS skeleton-only allocator.
At each window, I first run the test over the full set of allocators in both families. I then run over a narrowed selection of allocators. For the first family, this selection consists of the 4 allocators that cover the 2 entry points of information from the DAG
into HRP (Table~\ref{tab:crossing}) plus the second symmetrisation control, \undirhrp{} (where the distance matrix used for clustering only depends on which pairs are linked).
For the second family, the narrowed set consists of the 3 allocators from Table~\ref{tab:crossing} that are edge-orientation-aware.
I ran the narrowed tests after having obtained the results from the full sets of allocators. I therefore quote the full-set test results as my primary results. The $p$-values are in Table~\ref{tab:spa}, with the last column pooling all 4 windows into a single test against the 252-day baseline. This pooled test corrects for the search over windows as well as the search over allocators.

\begin{table}[t]
  \centering
  \caption{Hansen's SPA $p$-values by family, window and allocator set.}
  \label{tab:spa}
  \small
  \begin{tabular}{l l c c c c c}
    \toprule
    Family & Allocator set & 189 & 252 & 378 & 504 & Pooled \\
    \midrule
    (i) vs \corrhrp{}      & full     & $\rsSpaSkelWonePre$ & $\rsSpaSkelWtwoPre$ & $\rsSpaSkelWthreePre$ & $\rsSpaSkelWfivePre$ & $\rsSpaSkelPooledPre$ \\
    (i) vs \corrhrp{}      & narrowed & $\rsSpaSkelWone$ & $\rsSpaSkelWtwo$ & $\rsSpaSkelWthree$ & $\rsSpaSkelWfive$ & $\rsSpaSkelPooled$ \\
    (ii) vs \skelhrp{} & full     & $\rsSpaDirWonePre$ & $\rsSpaDirWtwoPre$ & $\rsSpaDirWthreePre$ & $\rsSpaDirWfivePre$ & $\rsSpaDirPooledPre$ \\
    (ii) vs \skelhrp{} & narrowed & $\rsSpaDirWone$ & $\rsSpaDirWtwo$ & $\rsSpaDirWthree$ & $\rsSpaDirWfive$ & $\rsSpaDirPooled$ \\
    \bottomrule
  \end{tabular}
  \floatnote{(i) denotes the larger family of all causal allocators, tested against \corrhrp{}. (ii) denotes the smaller family of only the edge-orientation-aware allocators, tested against \skelhrp{}. ``Pooled'' denotes SPA tests where the candidate set is the union of the family's allocators at all 4 windows against the baseline at the 252-day window. This accounts for both the strategy search and the window search. None of the entries reaches $p<0.05$.}
\end{table}

Over the full set of allocators, the null hypothesis is not rejected at any window for either of the two families.
Over the narrowed sets of allocators (the allocators that cover the two possible entry points of causal graphs into HRP), it is not rejected either. The first family never comes close ($p \geq \rsSpaSkelWfive$), which is what the sign counts of Section~\ref{sec:graph-vs-correlation} lead one to expect. The second family comes closest at the 504-day window ($p=\rsSpaDirWfive$), where the pairwise orientation effect also peaks.
The results in the pooled column show that, once the search across windows is accounted for as well, the $p$-values are $\rsSpaSkelPooled$ for the narrowed
causal family and $\rsSpaDirPooled$ for the narrowed orientation family.

Finally, the $90\%$ Model Confidence Set, the smallest set of strategies that contains the best one at the given confidence level, retains \rsMcsSize\ of the \rsMcsUniverse\ 252-day strategies and excludes only Hierarchical Sensitivity Parity as published (Appendix~\ref{app:hsp}).
With this many highly correlated variants, whose Sharpe ratios span $0.35$--$0.43$, the MCS cannot rank the family or separate treatments from controls.

\subsection{Out-of-sample testing}
\label{sec:oos}

To verify whether the in-sample results of the decomposition generalise, I run the same
DYNOTEARS allocators, and compare them to \corrhrp{}, on out-of-sample data: 19 rebalances from January 2025 to August 2026. 
I take price data from Yahoo Finance, as CRSP daily data ends on 2024-12-31. 
I find that the performance impact of the edge-orientation-aware allocators over the skeleton allocators is positive at all 4 windows for both \semcovhrp{} and for \toposemcovhrp{} (though at 189 days \toposemcovhrp{}'s orientation gain is only $+0.002$).
The total effect of the edge-orientation-aware allocators over \corrhrp{} is positive out of sample, from $+0.096$ to $+0.142$. This contrasts with the in-sample results that are statistically indistinguishable from zero at every window ($-0.017$ to $+0.013$, Table~\ref{tab:decomp}). This positive out-of-sample result is driven by the skeleton yielding a positive impact: in-sample, it had a negative effect relative to \corrhrp{}, but on this out-of-sample data, its effect is positive at 3 of the 4 windows (and is $-0.009$ at the 378-day window).
Thus, the skeleton's sign reversed out of sample, so its effect on performance is not stable across time.
The full out-of-sample results are located in Table~\ref{tab:oos}.
At the three longer estimation windows, the difference between the edge-orientation-aware allocators that use the DAG-implied covariance and the de-factored control (which clusters on the discovered causal skeleton, but uses a covariance built without the graph) matches the in-sample results: their Sharpe ratios are within $0.006$ of each other. 
However, at 189 days, \semcovhrp{} falls $0.090$ below the control, outside the in-sample band of roughly $\pm0.01$.
Note that, because the standard errors are too large on this small unseen sample, the point estimates above are sign checks rather than definitive.

\begin{table}[t]
  \centering
  \caption{Out-of-sample comparisons (net $\Delta$Sharpe, 2025-01-02 to 2026-08-07).}
  \label{tab:oos}
  \small
  \begin{tabular}{l c c c c}
    \toprule
    & \multicolumn{4}{c}{Window (days)} \\
    \cmidrule(lr){2-5}
    Contrast & 189 & 252 & 378 & 504 \\
    \midrule
    skeleton: \skelhrp{} $-$ \corrhrp{} & $\rsOosSkelWone$ & $\rsOosSkelWtwo$ & $\rsOosSkelWthree$ & $\rsOosSkelWfive$ \\
      & {\footnotesize($\rsOosSkelSeWone$)} & {\footnotesize($\rsOosSkelSeWtwo$)} & {\footnotesize($\rsOosSkelSeWthree$)} & {\footnotesize($\rsOosSkelSeWfive$)} \\
    \addlinespace[2pt]
    orientation: \semcovhrp{} $-$ \skelhrp{} & $\rsOosOrientWone$ & $\rsOosOrientWtwo$ & $\rsOosOrientWthree$ & $\rsOosOrientWfive$ \\
      & {\footnotesize($\rsOosOrientSeWone$)} & {\footnotesize($\rsOosOrientSeWtwo$)} & {\footnotesize($\rsOosOrientSeWthree$)} & {\footnotesize($\rsOosOrientSeWfive$)} \\
    \addlinespace[2pt]
    orientation: \toposemcovhrp{} $-$ \skelhrp{} & $\rsOosOrientBothWone$ & $\rsOosOrientBothWtwo$ & $\rsOosOrientBothWthree$ & $\rsOosOrientBothWfive$ \\
      & {\footnotesize($\rsOosOrientBothSeWone$)} & {\footnotesize($\rsOosOrientBothSeWtwo$)} & {\footnotesize($\rsOosOrientBothSeWthree$)} & {\footnotesize($\rsOosOrientBothSeWfive$)} \\
    \addlinespace[2pt]
    total: \semcovhrp{} $-$ \corrhrp{} & $\rsOosTotalWone$ & $\rsOosTotalWtwo$ & $\rsOosTotalWthree$ & $\rsOosTotalWfive$ \\
      & {\footnotesize($\rsOosTotalSeWone$)} & {\footnotesize($\rsOosTotalSeWtwo$)} & {\footnotesize($\rsOosTotalSeWthree$)} & {\footnotesize($\rsOosTotalSeWfive$)} \\
    \addlinespace[2pt]
    residual: \semcovhrp{} $-$ \dfhrp{} & $\rsOosResidWone$ & $\rsOosResidWtwo$ & $\rsOosResidWthree$ & $\rsOosResidWfive$ \\
      & {\footnotesize($\rsOosResidSeWone$)} & {\footnotesize($\rsOosResidSeWtwo$)} & {\footnotesize($\rsOosResidSeWthree$)} & {\footnotesize($\rsOosResidSeWfive$)} \\
    \bottomrule
  \end{tabular}
  \floatnote{Standard errors (in parentheses) are from the stationary block bootstrap of Section~\ref{sec:measurement} (21-day mean block size, 10{,}000 resamples), applied to the joint daily return panel of each pair.}
\end{table}

\subsection{Caveats}
\label{sec:caveats}

There are several things that could potentially undermine the validity of this study.

The first threat is low power: the effect sizes are smaller than the minimum detectable effect.
The difference in Sharpe ratio I observe is around $0.01$--$0.03$ over 18 years.
The corresponding block bootstrap standard errors at the 252-day window are $\rsContrastSETotal$ for the total contrast (\semcovhrp{}$-$\corrhrp{}) and $\rsContrastSEOrient$ for the orientation contrast, taking the larger of the two orientation contrasts (\toposemcovhrp{}$-$\skelhrp{}; the allocation-step contrast \semcovhrp{}$-$\skelhrp{} has a standard error of $0.012$), so the orientation bound below is the conservative one.
The small effect sizes relative to the standard errors indicate the study's lack of power.
At $5\%$ size and $80\%$ power, the one-sided minimum detectable effects
are ${\approx}\,\rsContrastMDETotal$ and
${\approx}\,\rsContrastMDEOrient$ net Sharpe respectively.
These minimum detectable effects are much larger than the observed orientation gains of $+0.009$ to $+0.021$ and skeleton costs of $-0.003$ to $-0.028$.
By the same token, none of the DYNOTEARS decomposition contrasts reaches conventional significance: the smallest $p$-value among the 12 orientation differences is $0.051$, which is approximately what multiplicity would predict under the null hypothesis anyway.
Some contrasts outside that set do reach $p<0.05$, but only prior to correcting for the search. Under VARLiNGAM, \skelhrp{}$-$\corrhrp{} at 189 and 252 days ($-0.035$ and $-0.033$, both $p=0.032$), \semcovhrp{}$-$\corrhrp{} at 189 and 252 days ($-0.071$, $p=0.036$, and $-0.042$, $p=0.046$), \toposemcovhrp{}$-$\corrhrp{} at 252 days ($-0.041$, $p=0.041$), and \coanchrp{}$-$\skelhrp{} at 504 days ($+0.030$, $p=0.012$) do. Under DYNOTEARS, so do the two contrasts involving the partial-correlation control: \pcorrhrp{}$-$\corrhrp{} at 189 days ($-0.039$, $p=0.008$) and \semcovhrp{}$-$\pcorrhrp{} at 189 days ($+0.035$, $p=0.017$). However, these are uncorrected $p$-values selected from a search over \rsNtrials\ configurations; the family-wise tests (Table~\ref{tab:spa}) reject nothing.
Rather than significance testing across different contrasts to establish an orientation effect, therefore, the main evidence for the existence of an orientation effect is the fact that there is a consistent positive contribution across windows and
treatments (Section~\ref{sec:decomposition}).

The second threat is that some of the design choices of my research were influenced by earlier results I had obtained.
The research question itself was informed by initial research that used Hierarchical Sensitivity Parity \cite{rodriguez2023hsp} (Section~\ref{sec:hierarchical-allocation}) as a baseline.
I correct for this in the Deflated Sharpe ratio: \rsNtrialsPhaseI\ of the \rsNtrials\ total configurations were from this initial research phase, with the remainder from the primary arm of the study.
The estimation windows themselves were also changed over time. I first used the conventional 252- and 504-day windows. These results indicated a temporal shift in the marginal contributions of the skeleton and the edge orientations; to analyse this further, I added in the 189- and 378-day windows.
However, the risk from these post hoc changes is low.
I ran causal discovery and allocation over the added windows identically to the initial windows.
Also, the total number of configurations, \rsNtrials, accounts for both the initial and subsequent (primary) research directions, across all phases and windows. The multiplicity-aware tests therefore account for the change in research direction.
The final design choice that was influenced by my earlier results is the family of allocators that I test.
Initially, I planned to test a family of 7 graph-consuming allocators. I then created a narrowed family to the 5 allocators after seeing the results of that larger family: it is the results of this narrowed family that I focus on in this section. The motivation for this narrower family is that it specifically covers the allocators where the causal graph enters HRP's two entry points. The two excluded allocators take in the graph elsewhere, and so are not directly relevant to the ablation design. Both are edge-orientation-aware allocators (Table~\ref{tab:allocator-map}): \structrp{} replaces the hierarchy with risk parity over structural shocks; \coanchrp{} constructs the clustering distance using shared upstream ancestry. 
Because this narrowing was post hoc, results from testing the 5-allocator family are indicative rather than definitive; the full 7-allocator family remains the primary SPA test.
Both families' SPA $p$-values are recorded in Table~\ref{tab:spa}.
The null hypothesis is never rejected for the full (7-allocator) family, nor for the narrowed family.
The results from the 2 allocators that were excluded from the narrowed family are directionally consistent with the rest of the allocators: they underperform the baseline in 12 of their 16 configurations, across both discovery methods and all 4 windows, by $-0.035$ to $+0.011$ net Sharpe. This is evidence that I did not drop allocators on the basis of their results alone.
Separately, I realised a potential objection lay in the notion that $\Sigma_{\mathrm{SEM}}$'s gain might be unrelated to causal direction, given that it also acts like shrinkage and assumes independent shocks and so de-factors assets. I therefore added the two mechanism controls discussed in Section~\ref{sec:orientation-role}.
I count them among the \rsNtrials\ configurations.

While all of the above changes were deliberate, one of my changes was a correction of an error I had made.
Initially, every allocator that conducts hierarchical clustering (including the \corrhrp{} baseline) projected its distance matrix onto the nearest positive semi-definite matrix before clustering. This step was an error that was inherited from the code from the first phase of my research; the projection collapsed the baseline's dendrogram into a list that was ordered based on a single scalar per asset.
Correcting this mistake raised \corrhrp{}'s net Sharpe by $+0.020$ to $+0.046$, while moving \skelhrp{} and \semcovhrp{} by a maximum of $0.005$. This turned my initial finding, that there was a performance gain from using the graph, into the null result I now report.
Note that the runs I conducted prior to this correction were the same configurations as I currently run, and did not consist of separate configurations. As a result, they do not add anything to the number of trials, \rsNtrials, so do not affect the Deflated Sharpe ratio or the SPA tests.

The third threat is the fact that I conducted this research on a very specific set of assets: an approximated version of the S\&P~100, using long-only allocation. I have not tested whether these results generalise to other asset classes. In addition, as mentioned in Section~\ref{sec:data}, my list of assets has selection bias, though this does not affect pairwise comparisons between allocators.

The fourth threat concerns the identification of edge orientations. DYNOTEARS
identifies them through its score and the acyclicity
constraint; Section~\ref{sec:causal-discovery} discusses the fact that, on unscaled data, this score
orients edges by marginal variance. I eliminate this effect by fitting on
standardised data. However, even after scaling, ordering by explained variance persists \cite{reisach2023r2}. For assets, explained variance is primarily due to market exposure, which implies that some portion of the discovered edge orientations may be based on how much market exposure each asset has, rather than a true causal ordering.
I did not assess this threat, but its magnitude could be calculated by estimating the fraction of edges that point from lower- to higher-$R^2$ assets. Another method would be to just use an ordering by $R^2$ alone (rather than a graphical ordering). This would represent a control for \topohrp{} in the same way that the partial-correlation
network controls for \skelhrp{}.
In any case, as all the allocators receive the same graph, the performance decomposition is unaffected.

One other caveat is the fact that my results are method-dependent, as the performance gain from edge orientations only exists under DYNOTEARS graphs. I therefore can only say that the orientation gain is specific to DYNOTEARS' contemporaneous structure, not make a broader claim about causal edge orientations being useful for portfolio allocation in general; and the de-factored control shows that even this gain cannot be attributed to edge direction itself.

Overall, the results are null. Replacing the correlation matrix in a hierarchical allocator with a DYNOTEARS graph does not improve performance at any estimation window: the total is within noise of zero everywhere, with negative point estimates at the two shorter windows and positive ones at the two longer, none of them significant. The graph's skeleton costs performance at every window, most at the 1-year window, and the edge orientations recover a comparable amount at every window, most at the 2-year window.
The point estimates of the two components have consistent signs across the estimation windows, but no contrast is significant after correcting for the search over allocators, whether or not the search over windows is also accounted for.
At the allocation step, a de-factored covariance (which does not use edge direction information) replicates the edge orientations' gain. This implies that the recovery seen in the edge-orientation-aware allocators cannot be attributed to the edge direction itself. Also, the recovery is only observed for the edge-orientation-aware allocators that use DYNOTEARS as a causal discovery method, not VARLiNGAM.

\section{Conclusion}
\label{chap:conclusion}
\label{sec:summary}

The research question this paper sought to answer was the following:
when the correlation matrix in
hierarchical portfolio allocation is replaced by a causal graph, does
performance improve, and what part of any improvement is attributable to
the graph's skeleton versus its edge orientations?
Two sub-questions posed in Section~\ref{sec:question} then addressed the mechanisms underpinning the performance of those two aspects of the graph:
whether the skeleton's effect is specific to the discovered DAG, or
reproducible by any comparably dense sparse network; and whether the
edge-orientation-aware allocators' effect comes from edge direction, or just from the fact that the SEM covariance involves shrinkage and the removal of market factors.
The ablation of Section~\ref{sec:ablation-design} answers all three of the above questions, and the statistical inference 
I conduct in Section~\ref{chap:evaluation} quantifies the results.

I find that replacing the correlation matrix with a causal graph does not improve
performance at any estimation window. Within this net-zero effect, the graph's skeleton has a negative performance effect, while the edge orientations offset this with a positive effect. None of the comparisons I conduct are statistically significant after correcting for the search over allocator configurations. I also find that the two decomposed performance effects do not even require a graph: a graph-free sparse network reproduces the skeleton's effect, while a direction-free de-factored covariance reproduces the edge orientation's effect.
I summarise the key results below:

\begin{enumerate}[itemsep=2pt]
  \item Using causal graphs yields no improvement over correlation matrices. 
        The DYNOTEARS graph has a relative effect of $-0.017$ to $+0.013$ net Sharpe
        (\semcovhrp{} $-$ \corrhrp{}) across the 4 windows, with none of those differences being significant ($p \geq 0.36$).
        Hansen's SPA test does not reject at any window, nor when the windows are pooled (Table~\ref{tab:spa}). The $90\%$ Model Confidence Set similarly does not reject its null hypothesis. Thus the answer to the main research question is no.
  \item The performance decomposition of the graph-fed allocators is consistent across windows, with the skeleton's effect being negative and the edge orientations' contribution being positive (at least when compared to \skelhrp{}) at every window. The two effects are of similar size, so they offset each other and the total effect is near zero. In any case, both components' effect sizes are below the minimum detectable effect (approximately \rsContrastMDEOrient{} Sharpe for the comparison between edge-orientation-aware and skeleton-only allocators), so effects below that threshold cannot be detected (Section~\ref{sec:caveats}). The same signs of the performance effects of both aspects of the graph are also observed under HERC.
  \item The negative performance effect of the skeleton does not require a discovered graph: a similarly dense
        partial-correlation network, built
        without discovery, reproduces the skeleton's negative performance compared to \corrhrp{} at
        every window ($p \geq 0.07$ for the difference). Thus, the discovered causal graph adds no additional information at the clustering 
        step that is not already contained in a generic sparse dependence network (Section~\ref{sec:skeleton-role}). This answers the first
        sub-question.
  \item The performance effect of using edge orientations is due to the removal of the common market factor, rather than directional information or shrinkage: a de-factored covariance, built without a graph, reproduces the
        effect at every window (residual $-0.011$ to $+0.001$) (Section~\ref{sec:orientation-role}). 
        Note that the edge orientations from VARLiNGAM graphs do not yield a performance improvement, instead clustering around zero. Using ridge-Granger graphs does not have a meaningful effect on the performance of the ordering-only allocator, while \semcovhrp{} falls, indicating that SEM-implied covariance requires reliable edge magnitudes, and not just directional information alone. This answers the second
        sub-question.
  \item In out-of-sample testing (January 2025 to August 2026, 19 rebalances,
        Section~\ref{sec:oos}), the effect of the edge orientations over the skeleton and the total effect of causal graphs over \corrhrp{} were positive.
\end{enumerate}

My findings align with the broadly mixed picture from prior work in Section~\ref{chap:background} on the application of causal networks to portfolio allocation.
Howard, Lohre and Mudde, though they never used causal graphs as inputs for allocators, found that causal networks could not consistently outperform correlation \cite{howard2025causal}. They also found that the contemporaneous matrix carries nearly all of the structure, which supports my decision to just use that matrix.
The network-portfolio literature found that under-weighting central nodes improves allocation because centrality is closely related to market exposure \cite{pozzi2013spread,peralta2016network}. The de-factored control that I use shows that the performance gain of edge-orientation-aware allocators is underpinned by the very same mechanism.

\subsection{Contributions}
\label{sec:contributions}

In Section~\ref{sec:challenges}, I set out three challenges, as well as three contributions that answered them respectively. I restate them below, in light of what my research found.

\begin{itemize}[itemsep=2pt]
  \item The first challenge is that of attribution: that
        the different aspects of a causal graph inform allocation jointly, and so
        cannot be individually isolated. My contribution is to set up and execute an ablation design
        (Section~\ref{sec:ablation-design} and
        Section~\ref{sec:decomposition}), as well as a discovery pipeline to supply the causal graph
        (Section~\ref{chap:discovery}). In each window, the same graph (from the DYNOTEARS
        or VARLiNGAM algorithm) feeds allocators. These different allocators take in incrementally
        more information from the DAG, so that each increment isolates one aspect of
        the graph. 
        In Sections~\ref{sec:orientation-role}
        and~\ref{sec:skeleton-role}, I apply mechanism controls to test the mechanisms driving the performance effects behind each of the components, and check whether the graph is necessary
        at all. 
        The pipeline is deterministic, so all results are reproducible. 
        This decomposition is a novel contribution: no prior work separately attributes the effect on allocation performance of the causal graph's skeleton from that of
        its edge orientations.
  \item The second challenge is that hierarchical
        allocators can only take in symmetric inputs, and so cannot use edge orientation information. My contribution is the creation of a novel edge-orientation-aware allocator family (Section~\ref{chap:allocators}): the closest prior work
        \cite{howard2025causal} never actually sets portfolio weights from the graphs it discovers. 
        I use the SEM-implied covariance, which by construction is edge-orientation-aware, to incorporate orientation information into the allocation step of hierarchical allocators.
        I also use causal-ordered bisection, which replaces the clustering
        dendrogram with the DAG's topological order, as an edge-orientation-aware input into the ordering step. Three allocators, \semcovhrp{}, \topohrp{}, and \toposemcovhrp{}, cover the above two entry points for directional information, as well as their combination.
  \item The third challenge is conducting statistical inference at realistic effect sizes, given differences in net Sharpe that are small relative to standard errors, and also while accounting for the search over \rsNtrials\ configurations. 
  My contribution in this regard is the evaluation (Section~\ref{chap:evaluation}) and its findings. I conduct comprehensive inference using block-bootstrap contrasts, a Deflated Sharpe ratio over all \rsNtrials\ configurations, Hansen's SPA over two different allocator families, the Model Confidence Set, the minimum detectable effect, and an out-of-sample check.
        The findings from my research represent a novel empirical contribution.
        Over 4 estimation windows, I find that the causal graph yields no
        total gain over the correlation matrix in hierarchical allocation. The skeleton alone reduces performance relative to the
        correlation matrix, while the edge orientations
        recover this negative performance. None of the comparisons are significant after accounting for the search over allocators. The
        mechanism controls I examine indicate that the performance effect attributed to the edge-orientation-aware allocators is due to the removal of the common market factor rather than edge direction, and also that a sparse network built without any causal discovery reproduces the skeleton's negative performance effect.
\end{itemize}

\subsection{Limitations}
\label{sec:limitations}

There are naturally several limitations of this work.
First, the performance effects from the skeleton and orientation components are worth at most $0.03$ Sharpe. 
This is close to the minimum detectable effect at this sample size: at 80\% power, these are ${\approx}\,\rsContrastMDETotal$ (total) and ${\approx}\,\rsContrastMDEOrient$ (orientation) net Sharpe, approximately twice the observed effects (see Section~\ref{sec:caveats}).
The result is also not significant after correcting for the search over different allocators, over either the full or the narrowed family, at any estimation window (or once the windows are pooled).
Second, the positive effect from using edge-orientation-aware allocators is only present with DYNOTEARS as the causal discovery method, so it seems to be more a result of DYNOTEARS' contemporaneous structure rather than of edge orientations from causal DAGs in general.
Also, the de-factored control shows that the performance gain from using edge orientations in the allocation step can be achieved without the need for a graph, and similarly, the partial-correlation control shows that the graph's skeleton's impact on performance is mirrored by a graph-free sparse network.
Third, my research only covers long-only US large-cap equities; my findings may not generalise to other assets.
Fourth, I selected the universe of 99 assets from the largest S\&P~500
constituents across three dates in 2018--2020, and held it fixed for the whole
2007--2024 sample. This induces survivorship bias (Section~\ref{sec:data}). This only affects absolute Sharpe ratios, rather than comparisons between allocators, because all allocators use the same assets.
Fifth, the SEM is only built from the asset block, relegating the impact of the macro drivers to the residuals, which therefore share a common driver component. This means that the assumption of independent structural shocks is violated.
Sixth, earlier results influenced three aspects of my research. The research
question itself was informed by the results of the first phase (the
HSP research). I also increased the number of estimation windows from 2 to 4 partway through the study. I also narrowed the family tested by SPA from the full 7 allocators to 5 after obtaining the results (Section~\ref{sec:caveats}). I do state these influences above, and the DSR corrects for the fact that I searched over \rsNtrials\ configurations in total. 
Seventh, at the two longer windows, there is a difference in net Sharpe between the two orientation-blind controls: \undirhrp{} $-$ \skelhrp{} $= +0.008$ at the 378-day window, $p=0.47$, and $= +0.019$ at the 504-day window, $p=0.072$. 
As a result, there is additional estimator uncertainty at the 378- and 504-day
windows. Against \undirhrp{}, the orientation contribution at those windows is close to zero, while the edge orientations' gain relative to the \skelhrp{} control is the largest in the decomposition at the 504-day window. 
Eighth, I have not examined whether the discovered edges persist between monthly rebalances.
Given that the DAG-fitting is deterministic and many of the aggregate metrics are stable, I believe it is likely that the graph itself is temporally stable, but I do not properly examine this by using, for example, an edge-overlap measure.
Ninth, I did not address the risk that, even on standardised data, DYNOTEARS may simply
orient edges from low- to high-$R^2$ assets, as discussed in Section~\ref{sec:caveats}. 
For equities, $R^2$ is mostly market exposure, so part of the informational content of the edge orientations may be due to ordering on market beta rather than causal direction.
Tenth, I did not test different values of the DYNOTEARS L1 regularisation coefficients: I kept them at a fixed value of $\lambda_W=\lambda_A=0.05$. Thus, the sensitivity of the decomposition to that penalisation parameter is unknown.

\subsection{Broader impact}
\label{sec:impact}

For investors, the relevance of this study is that research into portfolio allocation is used by asset managers and pension schemes. Savers bear the consequences, albeit diffusely and after a long time interval. 
As discussed in Section~\ref{sec:hierarchical-allocation}, one of the main problems in this literature is that backtests of single, overfit strategies can be convincing in isolation.
Better research practices can protect investors; in this case, conducting evaluations that correct for the search over many candidate strategies or configurations. I apply this to my research here. 
Another issue is ``causal-washing'': merely labelling something as ``causal'' invites more trust than may be warranted. In this research, I find that the additional value of a causal method can be zero or negative, with the causally-discovered graph's skeleton being a worse clustering input than the standard correlation matrix, and the edge orientations that counteract this loss only do so via a market-factor removal mechanism that does not require a graph at all.
Actually conducting this decomposition can thereby uncover the truth behind the label.

\subsection{Future work}
\label{sec:future}

There are several angles for future work.
First, given that method-dependence is the major caveat to decomposition, additional discovery methods that provide orientation information should be used.
Repeating the ablation design on graphs from a different family
would test whether the value of orientation information generalises beyond DYNOTEARS. One such candidate is
non-linear NTS-NOTEARS \cite{sun2023ntsnotears}, though it is very computationally intensive to fit and I was therefore unable to use it in this study.
Second, evaluate an alternative allocation strategy: equalising risk contributions across the structural shocks in the SEM. My research allocates across assets so as to equalise their contributions to portfolio risk (risk parity). However, future work could allocate based on where risk originates in the causal graph. \structrp{}, one of the ancillary allocators I use, implements a long-only version of this, but to truly achieve parity over shocks, a long--short setup would be required.
Third, use a wider variety of assets (including different classes of assets) beyond the 99 assets I consider in this research; this would be informative regarding whether my results generalise.
Fourth, use edge-orientation-aware risk attribution. Using the total effect matrix $T$ allows for the decomposition of the portfolio return into the constituent structural (original) shocks. This decomposition cannot be expressed by standard correlation matrices, and it also allows for better testing than the simple Sharpe comparisons I use, because real-world firm-specific shocks (such as news and earnings announcements) can be used to verify the edge directions.
Fifth, quantify the survivorship
bias of Section~\ref{sec:data} by only using the firms that were actually members of a given index like the S\&P~500 at each rebalance date, and including the final delisting returns where relevant. Testing this would then show whether my results still hold on this point-in-time panel.

\section*{Code and data availability}

The source code is available at
\url{https://github.com/j-muthu/portfolio-optimisation-causal-networks}.
The raw market data is licensed from WRDS/CRSP and cannot be distributed, so reproducing the reported numbers requires WRDS/CRSP access. Without it, the code falls back to Yahoo Finance prices, which lack delisted tickers and are adjusted differently, so they will not reproduce the reported numbers exactly. Appendix~\ref{app:repro} gives further reproduction details.

\section*{Acknowledgements}

I thank Ce Guo, Wayne Luk and Stephen Weston for their supervision. I used Anthropic's Claude models (Opus 4.6--5 and Fable 5) as coding assistants and as drafting and proofreading aids.

\printbibliography[heading=bibintoc, title={References}]

\clearpage
\appendix

\section{Supplementary figures and reproducibility}
\label{app:tables}

This appendix contains the full tables discussed in
Section~\ref{chap:evaluation}, as well as the reproducibility notes.

\subsection{The allocator map}
\label{app:allocator-map}

Table~\ref{tab:allocator-map} places every allocator evaluated in this
study on the two axes of Section~\ref{chap:allocators}: where the leaf
ordering comes from, and which covariance recursive bisection splits
on. The primary family of Table~\ref{tab:variants} occupies six of its
cells. The remaining cells are the mechanism controls of
Sections~\ref{sec:orientation-role} and~\ref{sec:skeleton-role}, the HERC
cells of Section~\ref{sec:robustness}, and the two members of the full family that were excluded from the narrowed family
(Section~\ref{sec:caveats}). The table also fixes the name used for
each of them in the text.

\subsection{The full allocator matrix}

The net annualised Sharpe ratios of all configurations evaluated as part of the ablation design are shown in 
Table~\ref{tab:matrix-full}. The table also includes the ancillary results, such as those of the two allocators that are excluded from
the narrowed 5-allocator family, \structrp{} and \coanchrp{}. 
Indeed, out of all of the VARLiNGAM-fed allocators, \coanchrp{} at the 252- and 378-day windows ($0.418$ and $0.419$) is the highest.
I include both of these allocators' configuration counts in the total count of \rsNtrials\ configurations, and they also feature in the wider family that I run the SPA test over.
As mentioned, the ridge-Granger control has a limited scope, covering only a select few allocators: \skelhrp{} for the skeleton, \semcovhrp{} for the magnitudes, \topohrp{} for the ordering, and \structrp{} for the SEM-implied covariance.
It is also only used at the 252-day window; as mentioned, I use it simply to separate the effect of causal ordering from edge magnitudes, rather than as another full discovery allocator.

I also show the mechanism controls and the
HERC cells at every window in Table~\ref{tab:controls-full}; their reference allocators are in Table~\ref{tab:matrix-full}.
All three mechanism controls are included in the \rsNtrials-configuration count.

\begin{table}[htbp]
  \centering
  \caption{Net annualised Sharpe of the mechanism controls and the HERC allocators at every window (DYNOTEARS graphs).}
  \label{tab:controls-full}
  \begin{tabular}{l l c c c c}
    \toprule
    & & \multicolumn{4}{c}{Window (days)} \\
    \cmidrule(lr){3-6}
    Allocator & Role & 189 & 252 & 378 & 504 \\
    \midrule
    \multicolumn{6}{l}{\emph{Covariance controls (Section~\ref{sec:orientation-role})}} \\
    \lwhrp{}     & shrinkage, no orientation   & $\rsDlwSharpeWone$ & $\rsDlwSharpe$ & $\rsDlwSharpeWthree$ & $\rsDlwSharpeWfive$ \\
    \dfhrp{}     & de-factored, no orientation & $\rsDdfSharpeWone$ & $\rsDdfSharpe$ & $\rsDdfSharpeWthree$ & $\rsDdfSharpeWfive$ \\
    \midrule
    \multicolumn{6}{l}{\emph{Graph-free skeleton control (Section~\ref{sec:skeleton-role})}} \\
    \pcorrhrp{}  & graph-free, density-matched & $\rsDpcSharpeWone$ & $\rsDpcSharpe$ & $\rsDpcSharpeWthree$ & $\rsDpcSharpeWfive$ \\
    \midrule
    \multicolumn{6}{l}{\emph{HERC arm (Section~\ref{sec:robustness})}} \\
    \corrherc{}   & baseline           & 0.454 & 0.410 & 0.345 & 0.364 \\
    \skelherc{}   & skeleton member    & 0.197 & 0.228 & 0.316 & 0.288 \\
    \semcovherc{} & orientation member & 0.285 & 0.280 & 0.326 & 0.337 \\
    \bottomrule
  \end{tabular}
  \floatnote{The covariance controls keep \skelhrp{}'s clustering step and only change the covariance that recursive bisection splits on. \pcorrhrp{} keeps the sample covariance and changes the skeleton. The HERC allocators use the same dendrograms and bisection covariances as \corrhrp{}, \skelhrp{}, and \semcovhrp{}, respectively.}
\end{table}

In Section~\ref{sec:robustness}, I use HERC as an alternative hierarchical allocator baseline. I find that using it as a baseline instead of HRP reproduces the signs of both comparisons at every window: the orientation comparison
(\semcovherc{} $-$ \skelherc{}) has a positive net Sharpe difference: $+0.088$,
$+0.052$, $+0.010$, $+0.049$ (unadjusted bootstrap $p=0.10$, $0.38$,
$0.90$, $0.46$), while the skeleton contrast (\skelherc{} $-$ \corrherc{}) is
negative: $-0.258$, $-0.182$, $-0.028$, $-0.076$ ($p \geq
0.07$).

\subsection{Headline configurations in return terms}

In Table~\ref{tab:return-terms}, I display several other metrics for the main allocator configurations at the 252-day estimation window: net CAGR, maximum drawdown, and
turnover.

\begin{table}[htbp]
  \centering
  \caption{Net CAGR, maximum drawdown, and mean one-way turnover per rebalance for the headline configurations at the 252-day window (DYNOTEARS graphs).}
  \label{tab:return-terms}
  \begin{tabular}{l c c c c}
    \toprule
    Variant & Net Sharpe & Net CAGR & Max drawdown & Turnover \\
    \midrule
    Equal weight ($1/N$) & 0.364 & 5.32\% & $-61.9$\% & 0.003 \\
    Inverse variance     & 0.388 & 5.21\% & $-51.0$\% & 0.027 \\
    \corrhrp{}       & 0.427 & 5.77\% & $-49.7$\% & 0.128 \\
    \skelhrp{}  & 0.399 & 5.37\% & $-50.2$\% & 0.118 \\
    \undirhrp{}   & 0.401 & 5.39\% & $-51.1$\% & 0.145 \\
    \semcovhrp{}   & 0.410 & 5.60\% & $-50.0$\% & 0.110 \\
    \toposemcovhrp{} & 0.406 & 5.52\% & $-50.6$\% & 0.140 \\
    \bottomrule
  \end{tabular}
\end{table}

\subsection{Sparsity and cost sweeps}

Table~\ref{tab:tau} reports the graph-sparsity sweep: net Sharpe on
DYNOTEARS graphs at the 252-day window as the edge-magnitude threshold
$\tau$ rises. Table~\ref{tab:cost} reports the transaction-cost sweep at
the 252-day and 504-day windows, with the mean one-way turnover of each
allocator in the final column.

\begin{table}[htbp]
  \centering
  \caption{Net Sharpe values against different edge-magnitude thresholds $\tau$ (DYNOTEARS, 252-day window).}
  \label{tab:tau}
  \begin{tabular}{l c c c c}
    \toprule
    Allocator & $\tau=0$ & $\tau=0.01$ & $\tau=0.05$ & $\tau=0.10$ \\
    \midrule
    \skelhrp{} (skeleton) & 0.399 & 0.399 & 0.386 & 0.395 \\
    \topohrp{}            & 0.399 & 0.399 & 0.403 & 0.402 \\
    \structrp{}           & 0.397 & 0.397 & 0.401 & 0.407 \\
    \bottomrule
  \end{tabular}
\end{table}

\begin{table}[htbp]
  \centering
  \caption{Net Sharpe by one-way cost at the 252-day and 504-day windows (DYNOTEARS).}
  \label{tab:cost}
  \begin{tabular}{l c c c c c}
    \toprule
    Allocator (window) & 0\,bps & 5\,bps & 10\,bps & 20\,bps & turnover \\
    \midrule
    \skelhrp{} (252)  & 0.403 & 0.399 & 0.394 & 0.386 & 0.118 \\
    \semcovhrp{} (252)  & 0.414 & 0.410 & 0.406 & 0.398 & 0.110 \\
    \toposemcovhrp{} (252) & 0.411 & 0.406 & 0.401 & 0.391 & 0.140 \\
    \skelhrp{} (504)  & 0.379 & 0.375 & 0.371 & 0.364 & 0.107 \\
    \semcovhrp{} (504)  & 0.399 & 0.396 & 0.393 & 0.386 & 0.089 \\
    \toposemcovhrp{} (504) & 0.403 & 0.399 & 0.395 & 0.386 & 0.120 \\
    \bottomrule
  \end{tabular}
  \floatnote{Turnover denotes the mean fraction of the portfolio traded per rebalance.}
\end{table}

\FloatBarrier

\subsection{Regime-conditional Sharpe (252-day window)}
\label{app:regimes}

\begin{figure}[tbp]
  \centering
  \includegraphics[width=0.9\textwidth]{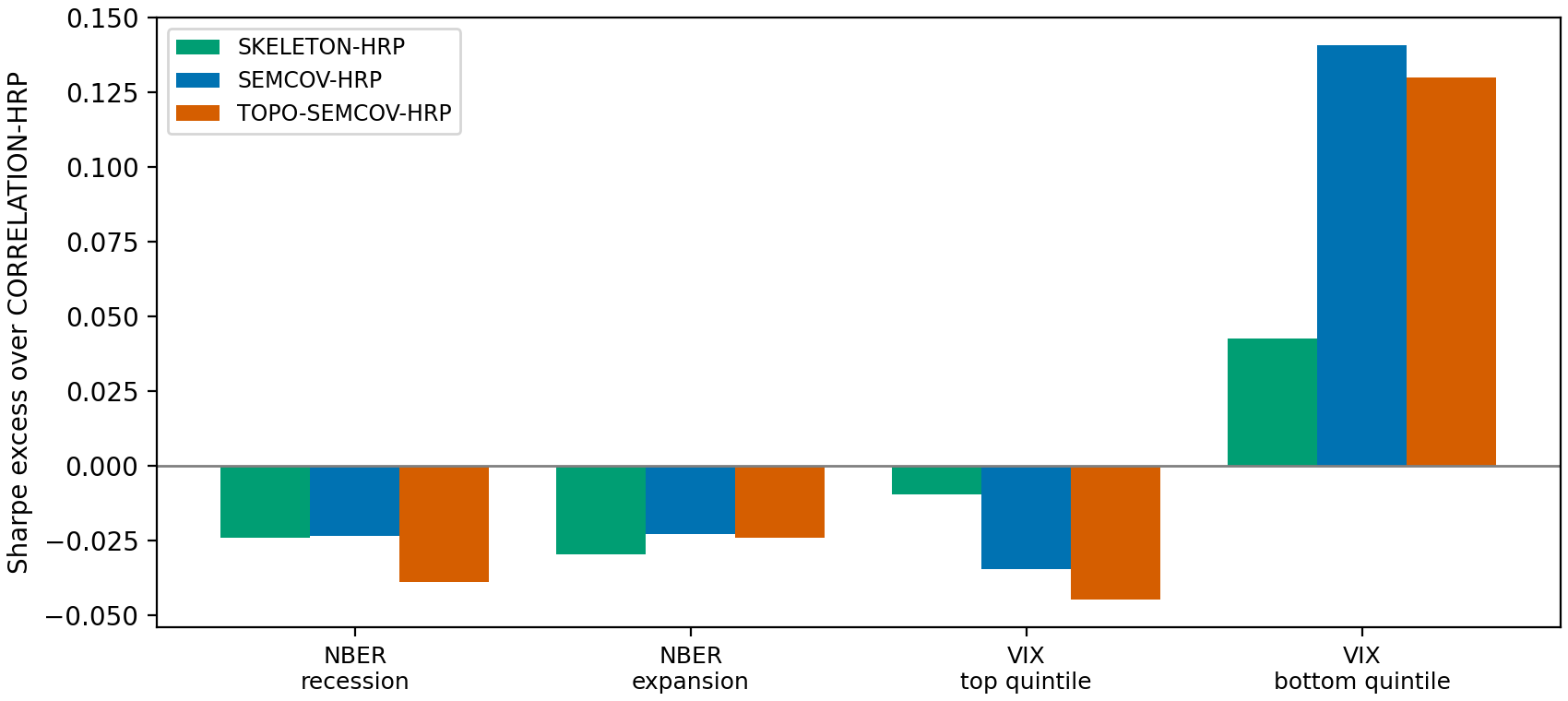}
  \caption{Regime-conditional excess Sharpe over the \corrhrp{} baseline (252-day window).}
  \label{fig:regime}
\end{figure}

As mentioned in Section~\ref{sec:causal-investing}, prior literature has reported that causal networks carry information about macro-financial regimes that correlation alone misses. This suggests that edge orientations may be most informative during periods of market stress, when sample moments are least reliable. 
To examine this hypothesis, I classify the daily returns of the 252-day-window portfolios based on whether they fall within NBER-classified recession and expansion dates and top and bottom quintiles of the VIX.
The excess Sharpe ratios over the \corrhrp{} baseline are shown in Figure~\ref{fig:regime} for \skelhrp{}, \semcovhrp{} and
\toposemcovhrp{}, and
Table~\ref{tab:regime-full} gives the net Sharpe ratios. 
I find that the point estimates contradict the hypothesis: 
in all regimes except the bottom VIX quintile, the alternative allocators exhibit worse performance than the \corrhrp{} baseline.

\begin{table}[htbp]
  \centering
  \caption{Regime-conditional annualised net Sharpe, 252-day window.}
  \label{tab:regime-full}
  \begin{tabular}{l c c c c c}
    \toprule
    Variant & All & NBER rec. & NBER exp. & High-vol & Low-vol \\
    \midrule
    \corrhrp{}     & 0.427 & $-0.567$ & 0.786 & $-1.316$ & 5.746 \\
    \skelhrp{} (skeleton)     & 0.399 & $-0.591$ & 0.756 & $-1.326$ & 5.789 \\
    \semcovhrp{}                & 0.410 & $-0.590$ & 0.763 & $-1.351$ & 5.887 \\
    \toposemcovhrp{}               & 0.406 & $-0.606$ & 0.762 & $-1.361$ & 5.876 \\
    \bottomrule
  \end{tabular}
  \floatnote{Conditional Sharpe levels are mechanically inflated in calm regimes, so only within-column comparisons are meaningful.}
\end{table}

\FloatBarrier

\subsection{Initial research direction: HSP}
\label{app:hsp}

The initial direction of this study was whether replacing Hierarchical Sensitivity Parity's (HSP; Section~\ref{sec:hierarchical-allocation}) correlation-based driver selection with selection from a discovered causal graph improved allocation. 
Following the methodology of Section~\ref{chap:evaluation}, I evaluated 3 configurations: standard HSP, a
variant that selected drivers based on their total influence on
assets within the discovered causal DAG, and a variant that added in a feedback loop to adjust the drivers' scores based on each month's excess Sharpe ratios.
I evaluated these over 252- and 504-day windows, and conducted a robustness check over different feedback parameter values and numbers of drivers. These different configurations account for \rsNtrialsPhaseI\ of the \rsNtrials\ total configurations that the Deflated Sharpe ratio corrects for.

\begin{figure}[tbp]
  \centering
  \includegraphics[width=0.85\textwidth]{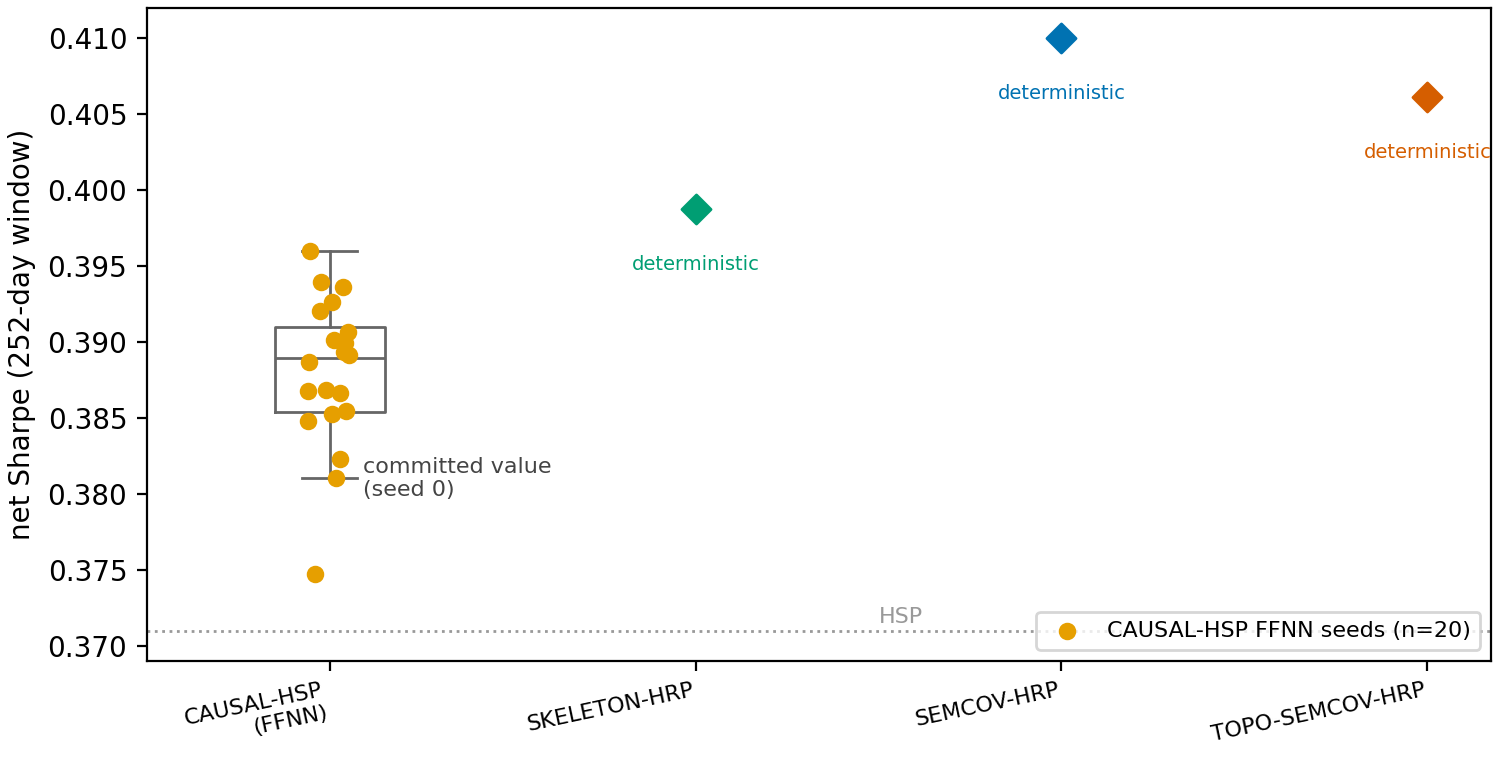}
  \caption{Net Sharpe ratio of the causal variant of HSP across \rsSeedN\ network seeds at the 252-day estimation window.}
  \label{fig:seed}
  \floatnote{The dotted line is the net Sharpe of standard HSP.}
\end{figure}

I found no detectable improvement from the causal graph.
At the 252-day window with 17 drivers, causal selection outperforms standard HSP
by $+0.010$ net Sharpe ($p=0.27$), but at other
driver counts, the sign reverses ($-0.015$, $-0.016$ and $-0.010$ at $K=10$, $20$ and
$25$). In any case, none of the differences is significant ($p=0.10$--$0.77$).
Running the same configuration with \rsSeedN\ different network seeds gives a range of net Sharpe ratios: $[\rsSeedMin, \rsSeedMax]$,
with a median of $\rsSeedMedian$ (Figure~\ref{fig:seed}). 
The seed range ($\rsSeedRange$) itself is therefore more than double the $+0.010$ performance gain from using the DAG in place of the correlation-based selection. 
In addition, standard HRP (\corrhrp{}) outperforms HSP at the same window ($\rsCorrSharpe$ against $0.371$).

I also find that adding the monthly feedback loop has no effect: across the full parameter grid, the closed-loop weights equal the open-loop weights at every rebalance.
This is because the assignment function I use can only re-rank drivers within the
causally selected set, and the mean monthly feedback has a very low signal-to-noise ratio of
approximately $\rsRewardSNR$ per update.

\subsection{Universe and driver lists}
\label{app:universe}

The asset universe contains 99 names: AAPL, ABBV, ABT, ACN, ADBE, ADP,
AEP, AMGN, AMT, AMZN, AXP, BA, BAC, BKNG, BLK, BMY, C, CAT, CHTR, CI, CL,
CMCSA, COF, COP, COST, CRM, CSCO, CVS, CVX, D, DE, DHR, DIS, DUK, EBAY,
ED, EOG, ETN, EXC, F, FDX, GE, GILD, GIS, GM, GOOGL, GS, HD, HON, IBM,
INTC, JNJ, JPM, KMB, KO, LIN, LLY, LMT, LOW, MA, MCD, MDT, META, MMM, MO,
MRK, MS, MSFT, NEE, NKE, NVDA, ORCL, PEP, PFE, PG, PM, PNC, PSX, RTX,
SBUX, SCHW, SLB, SO, SPG, SRE, T, TFC, TGT, TMO, TMUS, TXN, UNH, UPS,
USB, V, VZ, WFC, WMT, XOM (tickers as of the CRSP mapping of
Section~\ref{sec:data}).

I use 33 series as drivers. These are: year-on-year CPI, core CPI,
industrial production, retail sales and housing starts;
unemployment-rate change; and Michigan sentiment change. I use the following rates and spreads: 3-month, 2-, 5-, 10- and
30-year Treasury yield changes, 10y$-$2y and 10y$-$3m slope changes, and
Baa$-$Aaa and Baa$-$10y credit-spread changes. For currencies, I use the dollar-index
change and the EURUSD, JPYUSD, GBPUSD, and CNYUSD log returns. For commodities, I use the WTI,
Brent, gold, natural gas, silver, and copper log returns. For volatility and
international equity, I use the VIX level and the EFA, EEM, EWJ, EWG, and EWU log returns.

\subsection{Hyperparameters}
\label{app:hyper}

\begin{itemize}
  \item Discovery
  \begin{itemize}
    \item DYNOTEARS: lag $p=1$, $\lambda_W=\lambda_A=0.05$, minimum edge threshold $0.01$ (the extraction threshold $\tau$ of
      Section~\ref{sec:asset-graph} therefore defaults to $0$ and prunes
      nothing further).
    \item VARLiNGAM: lingam pruning disabled.
    \item Ridge-Granger: ridge coefficient $\lambda=10^{-2}$, scaled by the length of the estimation window.
  \end{itemize}
  \item Allocators
  \begin{itemize}
    \item Single-linkage where clustering is used. The clustering
      distance is passed to linkage with no projection.
    \item $\Sigma_{\mathrm{SEM}}$: diagonal shock variances from fit-window
      residuals, nearest-PSD projection and ridge coefficient
      $10^{-6}\operatorname{tr}(\Sigma)/N$.
    \item Ties are broken in topological ordering first by total downstream influence, then by name.
  \end{itemize}
  \item Protocol
  \begin{itemize}
    \item The transaction cost is set as 5\,bps one-way (swept $\{0,5,10,20\}$).
    \item Windows: 189/252/378/504 trading days.
    \item Time between each rebalance: 21 trading days.
    \item 215 rebalances (2007-01--2024-11) at each of the 4 estimation windows. At the 378-day window and the 504-day window, the first 6 and 12 rebalances, respectively, use equal weighting of assets as there is not enough data to allow for hierarchical allocation.
    \item Bootstrap: Politis--Romano, 10{,}000 resamples, mean block length of
      21 trading days.
  \end{itemize}
\end{itemize}

\subsection{Reproducibility}
\label{app:repro}

The code for this study is at\\
\url{https://github.com/j-muthu/portfolio-optimisation-causal-networks}.\\
The code allows for the reproduction of all
the allocator results in Section~\ref{chap:evaluation}, given WRDS/CRSP access.
Note that the first-phase results on HSP (see Appendix~\ref{app:hsp}) are not
bit-reproducible due to randomness in the initialisation of the neural networks. However, I discuss a seed audit in that appendix to quantify the spread of those results.

The asset price data is from CRSP via WRDS, which requires a licence and a
WRDS credential. I downloaded the macro drivers from FRED and Yahoo Finance, and the 99-asset ticker list is committed to the repository as \path{scripts/phase_i_universe.txt}.
Without WRDS data, the code falls back to prices from Yahoo Finance; the resulting Sharpe ratios may differ.
I use Yahoo Finance prices for the out-of-sample slice of 2025--26 (Section~\ref{sec:oos}), since CRSP daily data ends on 2024-12-31.
When the first backtest is run, it fills a local cache of prices, drivers and graph fits, so no subsequent scripts need to make a network call. This also saves a lot of time, because each graph fit takes around 4 minutes per window at $d\approx132$.

I do not commit the cache to the repository due to its size and the fact that it contains licensed data.
However, I have committed the collated CSVs under \path{results/}, figures, and the generated macros that feed into the paper.

I use Python 3.13 in a virtual environment built from the
repository's requirements file. To avoid dependency conflicts, \texttt{causalnex}, \texttt{lingam} and \texttt{nts-notears}
are vendored folders that I took from GitHub as opposed to pip packages, so are contained within the repository.
To run the full reproduction end-to-end, \path{reproduction.py}, located at the repository root, will run all of the necessary scripts, from the unit tests to the compilation of the paper.

\clearpage
\begin{landscape}
\begin{center}
  \captionof{table}{All allocators evaluated, arranged by ordering and bisection covariance sources.}
    \label{tab:allocator-map}
  \small
  \setlength{\tabcolsep}{3.5pt}
  \setlength{\aboverulesep}{0pt}
  \setlength{\belowrulesep}{0pt}
  \setlength{\extrarowheight}{2pt}
  \begin{tabular}{l | l l l l | l}
    \toprule
    \diagbox[height=2.4\line]{Ordering from}{Bisects on}
      & sample $\Sigma$ & Ledoit--Wolf & de-factored & $\Sigma_{\mathrm{SEM}}$ & Graph content used \\
    \midrule
    \multicolumn{6}{l}{\emph{Graph-blind}} \\
    correlation dendrogram      & \shortstack[l]{\corrhrp{} or\\\corrherc{}} & & & & none (baseline) \\
    \midrule
    \multicolumn{6}{l}{\emph{Skeleton only}} \\
    skeleton-embedding dendrogram   & \shortstack[l]{\skelhrp{} or\\\skelherc{}} & \lwhrp{}$^{a}$ & \dfhrp{}$^{a}$ & \shortstack[l]{\semcovhrp{} or\\\semcovherc{}} & skeleton; $\Sigma_{\mathrm{SEM}}$ adds orientation \\
    undirected-adjacency dendrogram & \undirhrp{} & & & & skeleton (control on the control) \\
    partial-correlation dendrogram  & \pcorrhrp{}$^{a}$ & & & & edge count only (graph-free) \\
    \midrule
    \multicolumn{6}{l}{\emph{Edge-orientation-aware}} \\
    topological order           & \topohrp{} & & & \toposemcovhrp{} & orientation (ordering step) \\
    co-ancestry dendrogram      & \coanchrp{}$^{b}$ & & & & orientation (clustering step) \\
    none (risk parity)          & & & & \structrp{}$^{b}$ & orientation (covariance only) \\
    \bottomrule
  \end{tabular}
  \floatnote{Rows denote the source of the asset ordering; columns denote the covariance it uses for cluster variances. Where an entry lists two allocators, they are the HRP and HERC equivalents. $^{a}$: mechanism control added after the grid was seen; sees no edge direction. $^{b}$: in the pre-specified 7-allocator family, but not in the narrowed 5-allocator family.}
\end{center}
\end{landscape}

\begin{landscape}
\begin{center}
  \captionof{table}{Net annualised Sharpe, 2007--2024, by graph source, allocator and window.}
    \label{tab:matrix-full}
  \begin{tabular}{l l c c c c c c c}
    \toprule
    Graph & \shortstack{Window\\(days)} & \skelhrp{} & \undirhrp{} & \semcovhrp{} & \topohrp{} & \toposemcovhrp{} & \structrp{} & \coanchrp{} \\
    \midrule
    DYNOTEARS & 189 & 0.392 & 0.389 & 0.407 & 0.392 & 0.408 & 0.400 & 0.410 \\
    DYNOTEARS & 252 & 0.399 & 0.401 & 0.410 & 0.399 & 0.406 & 0.397 & 0.396 \\
    DYNOTEARS & 378 & 0.404 & 0.412 & 0.413 & 0.405 & 0.419 & 0.400 & 0.403 \\
    DYNOTEARS & 504 & 0.375 & 0.394 & 0.396 & 0.382 & 0.399 & 0.387 & 0.390 \\
    VARLiNGAM & 189 & 0.376 & 0.387 & 0.340 & 0.387 & 0.394 & 0.400 & 0.398 \\
    VARLiNGAM & 252 & 0.394 & 0.392 & 0.385 & 0.388 & 0.386 & 0.392 & 0.418 \\
    VARLiNGAM & 378 & 0.399 & 0.400 & 0.410 & 0.398 & 0.399 & 0.394 & 0.419 \\
    VARLiNGAM & 504 & 0.364 & 0.377 & 0.371 & 0.372 & 0.375 & 0.375 & 0.394 \\
    ridge-Granger   & 252 & 0.386 & --- & 0.352 & 0.393 & --- & 0.350 & --- \\
    \bottomrule
  \end{tabular}
  \floatnote{The graph-blind \corrhrp{} baseline has a net annualised Sharpe of $\rsCorrSharpeWone$, $\rsCorrSharpe$, $\rsCorrSharpeWthree$ and $\rsCorrSharpeWfive$ at the 189-, 252-, 378- and 504-day windows, respectively. ``---'': ridge-Granger was not used as an input to this allocator.}
\end{center}
\end{landscape}

\typeout{get arXiv to do 4 passes: Label(s) may have changed. Rerun}
\end{document}